\documentclass[%
 reprint,
 superscriptaddress,
 nofootinbib,
 amsmath,amssymb,
 aps,
 prx,
]{revtex4-2}

\usepackage{mytikz}
\usepackage[mode=buildnew]{standalone}
\usepackage[caption=false]{subfig}

\usepackage[hidelinks]{hyperref}
\usepackage{graphicx}
\usepackage{array}[=2016-10-06] 
\usepackage{dcolumn}
\usepackage{bm}
\usepackage{booktabs}
\usepackage{amsmath,amssymb,amsbsy,amstext, amsthm, amsfonts}
\usepackage{color}
\usepackage{slashed}
\usepackage{dsfont}
\usepackage{verbatim}
\usepackage[utf8]{inputenc} 
\usepackage{ragged2e}
\usepackage{mathtools}
\usepackage{xfrac}
\usepackage{blkarray}
\usepackage{youngtab}
\usepackage{adjustbox}
\usepackage{xcolor}
\definecolor{Burgundy}{RGB}{144,0,32}
\definecolor{Burgundy1}{RGB}{128,0,32}
\definecolor{Burgundy2}{RGB}{158,5,8}
\definecolor{VividBurgundy}{RGB}{159,29,53}

\usepackage{hyperref}
\hypersetup{
    colorlinks=true,
    citecolor=VividBurgundy,
    filecolor=black,
    linkcolor=black,
    urlcolor=VividBurgundy
}

\usepackage{tikz}
\usetikzlibrary{shapes,arrows,cd,chains,decorations.markings,decorations.pathmorphing,calc}
\tikzset{
->-/.style args={#1rotate#2}{decoration={markings, mark=at position #1 with {\arrow[scale=1.5,rotate = #2 ]{stealth}}}, postaction={decorate}}
}
\tikzset{curve/.style={settings={#1},to path={(\tikztostart)
    .. controls ($(\tikztostart)!\pv{pos}!(\tikztotarget)!\pv{height}!270:(\tikztotarget)$)
    and ($(\tikztostart)!1-\pv{pos}!(\tikztotarget)!\pv{height}!270:(\tikztotarget)$)
    .. (\tikztotarget)\tikztonodes}},
    settings/.code={\tikzset{quiver/.cd,#1}
        \def\pv##1{\pgfkeysvalueof{/tikz/quiver/##1}}},
    quiver/.cd,pos/.initial=0.35,height/.initial=0}
\tikzset{tail reversed/.code={\pgfsetarrowsstart{tikzcd to}}}
\tikzset{2tail/.code={\pgfsetarrowsstart{Implies[reversed]}}}
\tikzset{2tail reversed/.code={\pgfsetarrowsstart{Implies}}}

\def\Vhrulefill{\leavevmode\leaders\hrule height 0.7ex depth \dimexpr0.4pt-0.7ex\hfill\kern0pt}

\begin{document}

\title{Higgsing Transitions from Topological Field Theory\\ \& Non-Invertible Symmetry in Chern-Simons Matter Theories }

\author{Clay C\'ordova}
\email{clayc@uchicago.edu}
\affiliation{Kadanoff Center for Theoretical Physics \& Enrico Fermi Institute, University of Chicago}
\author{Diego Garc\'ia-Sep\'ulveda}
\email{dgarciasepulveda@uchicago.edu}
\affiliation{Kadanoff Center for Theoretical Physics \& Enrico Fermi Institute, University of Chicago}
\author{Kantaro Ohmori}
\email{kantaro@hep-th.phys.s.u-tokyo.ac.jp}
\affiliation{Faculty of Science, University of Tokyo, Japan}

\date{\today}

\begin{abstract}

\noindent Non-invertible one‐form symmetries are naturally realized in (2+1)d topological quantum field theories. In this work, we consider the potential realization of such symmetries in (2+1)d conformal field theories, investigating whether gapless systems can exhibit similar symmetry structures. To that end, we discuss transitions between topological field theories in (2+1)d which are driven by the Higgs mechanism in Chern-Simons matter theories. Such transitions can be modeled mesoscopically by filling spacetime with a lattice-shaped domain wall network separating the two topological phases.  Along the domain walls are coset conformal field theories describing gapless chiral modes trapped by a locally vanishing scalar mass.  In this presentation, the one-form symmetries of the transition point can be deduced by using anyon condensation to track lines through the domain wall network.  Using this framework, we discuss a variety of concrete examples of non-invertible one-form symmetry in fixed-point theories.  For instance, $SU(k)_2$ Chern-Simons theory coupled to a scalar in the symmetric tensor representation produces a transition from an $SU(k)_2$ phase to an $SO(k)_4$ phase and has non-invertible one-form symmetry $PSU(2)_{-k}$ at the fixed point.  We also discuss theories with $Spin(2N)$ and $E_7$ gauge groups manifesting other patterns of non-invertible one-form symmetry.  In many of our examples, the non-invertible one-form symmetry is not a modular invariant TQFT on its own and thus is an intrinsic part of the fixed-point dynamics.

\end{abstract}

\maketitle

\makeatletter
\def\l@subsubsection#1#2{}
\makeatother
\tableofcontents


\section{Introduction}

Gapped phases in (2+1) dimensions are characterized by a spectrum of massive anyons and patterns of non-abelian statistics.  In the continuum limit, these systems are described by topological quantum field theory (TQFT), a paradigm for mathematically organizing the resulting long-range correlations and quantum entanglement. While the system remains gapped, the long-distance physics is robust.  However, when the gap closes the system may transition from one topological phase to another.  Such transitions may be either first order, or more interestingly, gapless, and characterized by a conformal fixed point which is in general strongly coupled.  

One rich source of such transitions are Chern-Simons matter gauge theories.  In these models the transition is driven by a charged scalar field $\phi$ which is massive on one side of the transition and condensed on the other.  Such Chern-Simons matter theories have been intensively investigated and can serve as continuum models for deconfined quantum criticality \cite{Wang:2017txt}. (See \cite{Senthil:2018cru} for a review).  They are also a fruitful playground to explore dualities where the fixed point is described in two complementary fashions using distinct ultraviolet degrees of freedom \cite{Giombi:2011kc, Jain:2013gza, Aharony:2015mjs,Seiberg:2016gmd,Hsin:2016blu}.

One of our main goals in this work is to elucidate aspects of these transitions between topological phases from the vantage point of the infrared topological field theories that they connect.  To carry this out, we introduce a mesoscopic model of the transition built from a network of domain walls separating the two infrared phases.  The network is shaped as a lattice permeating spacetime (or space in a related Hamiltonian model).  Along each domain wall there are gapless (1+1)d chiral degrees of freedom, defined by a coset conformal field theory / gauged WZW model \cite{DiFrancesco:1997nk, GOODARD198588, Karabali:1988au, Gawedzki:1988hq}.  As we cross the domain wall, the potential for the scalar field $\phi$ changes shape: on one side $\phi$ is massive, and on the other side it develops a vacuum expectation value.  The (1+1)d fields on the domain walls may thus be viewed as edge modes confined to loci where the $\phi$  has vanishing mass. 

As parameters of the domain wall network are changed, our model transitions between the given topological phases and plausibly has a phase transition in the same universality class as the desired Chern-Simons matter fixed point.  As described, our model is conceptually similar to the coupled wire constructions of \cite{kane2002fractional} and related models \cite{TeoKane2014,Mong2014,HuKane2018}.  However, our approach is distinctly rooted in continuum field theory and does not, for instance, provide a microscopic lattice description of either the topological phases or the gapless domain wall modes.  Related constructions based on continuum field theory ideas have also appeared in \cite{Sopenko:2023utk}. 

\subsection{Topological Lines at Critical Points}

From the vantage point of the microscopic Chern-Simons gauge theory many symmetries of the fixed point are emergent, meaning they are valid only at the infrared universality class characterizing the transition.  This is a well-studied phenomenon for ordinary (zero-form) global symmetries where key examples of duality are characterized by emergent time reversal, or non-abelian global symmetry \cite{Son:2015xqa, Metlitski:2015eka, Seiberg:2016gmd, Cordova:2017kue}.  By contrast, the possibility of emergent one-form symmetries is much less explored.  A one-form symmetry is by definition, an extended line operator whose correlation functions depend only topologically on the support of the line \cite{Gaiotto:2014kfa}.  Such a line is an emergent symmetry when this topological dependence is a property only of the correlation functions of the fixed-point universality class, but not the microscopic gauge theory Lagrangian.  Note that while topological lines are ubiquitous in the gapped topological phases discussed above, their appearance at a gapless transition is more striking. 

The simplest possible notion of a one-form global symmetry is an abelian symmetry group.  For instance, these are common in abelian gauge theory and often occur as center symmetries of non-abelian gauge theory with suitable non-minimal matter \cite{Gaiotto:2014kfa}. In gapped phases, these abelian one-form symmetries are described by abelian anyons with characteristic fusion rules. Notice that such symmetries always act on Hilbert space by invertible (unitary) operators.  More exotic, and the focus of much of our discussion below, is the possibility of non-invertible one-form symmetry at a fixed point.  These operators form a braided fusion category and, while they commute with the Hamiltonian, they do not act by invertible (unitary) operators on Hilbert space. One of the main results of our analysis is to provide qualitatively new examples of Chern-Simons matter fixed points with precisely this form of novel emergent symmetry. Again we emphasize that while non-invertible one-form symmetries are characteristic in topological phases, described by the non-abelian anyons of the theory, their appearance in a gapless phase is a new phenomenon.

To explain more precisely what is new about the examples to follow, it is helpful to recall that given any (2+1)-dimensional theory with a finite non-anomalous zero-form global symmetry $K$, gauging $K$ produces a new theory with one-form symmetry given by the fusion category $\mathrm{Rep}(K)$.  Moreover, if $K$ is non-abelian, the fusion ring of $\mathrm{Rep}(K)$ is not described by any abelian group.  Thus, any such example gives a simple construction of non-invertible one-form symmetry.  In a sense then, examples of non-invertible one-form symmetries abound and indeed examples based on finite, or more generally, disconnected gauge groups have been discussed in the literature. See e.g.\ \cite{Nguyen:2021yld, Heidenreich:2021xpr, Kaidi:2021xfk,Cordova:2022rer, Bhardwaj:2022yxj,Bhardwaj:2022lsg, Bhardwaj:2025jtf}.  However, unfortunately one-form symmetries manufactured by this construction do not provide any significant dynamical insight.   Indeed, gauging the finite subgroup $K$ is a topological operation on a field theory and hence commutes with the renormalization group flow. Therefore, the presence of the $\text{Rep}(K)$ symmetry is simply an avatar $K$ symmetry of the related ``ungauged" theory.  In contrast with the above, the examples we discuss below will have non-invertible one-form symmetry but \emph{connected} gauge groups. In these models the presence of non-invertible one-form symmetry is a fundamentally new constraint on the dynamics at the fixed point. (A gapless example in (3+1)d  with a connected gauge group and non-invertible one-form symmetry is a model of axions \cite{Choi:2022fgx}.)

\subsection{Condensation, Higgsing, and Hierarchies}

To argue for the presence of these new symmetries we make use of the mesoscopic model described above.  Indeed, one of the key advantages of this model is that it manifestly possess the non-invertible symmetries throughout its phase diagram.  Therefore, the bulk of our technical analysis is to understand the precise one-form symmetries of the mesoscopic model.  As we explain, the key idea is to ask how an anyon worldline in one topological phase can penetrate the gapless domain wall and continue as a new worldline in a different gapped phase.  This in turn is a question of how to describe the Higgsed phase  of the gauge theory ($\phi$ condensed) in terms of the anyons of the massive phase of the gauge theory ($\phi$ not condensed).  

We achieve such a description by modifying the vacuum of the massive phase to include loops of the $\phi$ field viewed as anyons in the topological phase.  This procedure is complicated by two important physical effects:
\begin{itemize}
    \item The anyons described by $\phi$ quanta in general carry spin.  Hence, on their own, they cannot be consistently summed.
    \item When these anyons collide, they produce new anyons which must also be summed by consistency.
\end{itemize}
Both of these problems are solved in the mesoscopic model.  In that case, gapless degrees of freedom on the domain wall can be viewed as an edge state for a gapped bulk known as the coset TQFT.  By pairing with degrees of freedom from the coset TQFT, the $\phi$ quanta become effectively bosonic and may then condense.  This is technically achieved using anyon condensation \cite{Moore:1989yh, Frohlich:2003hm, Bais:2008ni, Kong:2013aya, Burnell:2017otf, Cong:2017ffh} which also explains how to consistently incorporate the fusion products.  Thus in our model, the Chern-Simons matter transition driven by the Higgs mechanism, itself a form of ``condensation," modifies the vacuum via anyon condensation in a product TQFT modeling both the massive phase and the coset. 

We remark that analogous considerations have been applied in the context of hierarchy constructions of fractional Hall states \cite{HierarchyHaldane,HierarchyHalperin} (See \cite{HierarchyReview} for a review).  In particular in \cite{zhang2024hierarchy} a transition from one fractional Hall state to another is modeled as anyon condensation in a product theory, which includes the initial state as well a new topological degrees of freedom.  Applying the logic of our mesoscopic construction, we anticipate that these additional topological degrees of freedom have as an edge mode, the chiral gapless fields that appear as the magnetic field is modulated.  One can then envision a mesoscopic domain wall network model of fractional Hall state transitions directly analogous to our construction below for Higgsing transitions.

\subsection{Implications of Higher Symmetry}

As one of our main results is the appearance of non-invertible one-form symmetries in Chern-Simons matter fixed points, we summarize here several of their key physical consequences.  Some of these features are common to any one-form symmetry, while others depend particularly on the fact that the symmetries are non-invertible. 
\begin{itemize}
    \item The non-invertible symmetry does not act on local operators in the fixed point theory \cite{Gaiotto:2014kfa}.  Therefore, it cannot be explicitly broken by any relevant deformation of the fixed point and must be present in all resulting phases.  In our examples this symmetry will be spontaneously broken in the gapped phases that resulted from the fixed point by relevant deformation.   
    \item The non-invertible one-form symmetry may form a modular invariant full TQFT on its own, in which case it plausibly decouples from the rest of the theory.  Alternatively, it may be non-modular in which case it cannot decouple from the fixed-point dynamics \cite{Mueger:2003ModularCategories,Mueger:2003QuantumDouble,drinfeld2010braided, Hsin:2018vcg}.  In the latter case, we deduce that the fixed point, if gapless, must support conformally invariant line-defects that are charged under this non-invertible one-form symmetry.\footnote{Here by line defect we mean an extended operator whose total dimension in spacetime is one.  Such an object may act as an operator at a fixed time, or alternatively may define a point-like defect in space for all time.} Upon relevant deformation to a gapped phase, these conformally invariant defects flow to the additional anyons needed to make the braiding matrix non-degenerate.
    \item The braiding of non-invertible one-form symmetries is an anomaly that must be reproduced in any phase resulting from the fixed point \cite{Gaiotto:2014kfa, Putrov:2025xmw}. When the fusion rules of the symmetry are not those of the representation ring of a finite group, the braiding must be non-trivial \cite{Deligne:TensorCategories, nlab:deligne's_theorem_on_tensor_categories}. Hence, such a symmetry is always anomalous and obstructs a trivially gapped phase.  
    \item Beyond its action on the vacuum, the one-form symmetry will organize the non-zero energy states into multiplets.  For instance, the spectrum of the theory with periodic boundary conditions is constrained by the presence of a one-form symmetry. When this symmetry is non-invertible the multiplets will be representations of algebras, not groups.  See \cite{Lin:2022dhv, Bartsch:2022mpm, Bartsch:2022ytj, Bhardwaj:2023ayw, Bartsch:2023wvv, Cordova:2024iti, Choi:2024tri} for related discussion.
\end{itemize}
We hope to explore many of these features in future work. 

\section{The Higgsing Transition } \label{Section2}

In this section, we motivate our study of Chern-Simons matter theories.  Our goal is to analyze the Higgsing transition using tools of topological quantum field theory and (1+1)d conformal field theory.  We introduce a mesoscopic model for this transition based on domain wall networks.  Using this model we provide a framework to find the topological lines, in general non-invertible, that exist at these fixed points. 

\subsection{Chern-Simons Matter Flows}

We consider a (2+1)d Chern-Simons theory coupled to charged scalar matter.  The defining data of such a field theory are given as follows:
\begin{itemize}
    \item A gauge group $G,$  and a Chern-Simons level $k\in\mathbb{Z}$.  We denote this pair as $G_{k}.$  (Below we take $G$ to be a connected compact Lie group.)
    \item A scalar field in a representation $\mathbf{R}$ of $G$.  (Below, we take $\mathbf{R}$ to be irreducible for simplicity.)  In general, the representation, and hence scalar is complex unless otherwise noted.  We denote the scalar as $\phi_{\mathbf{R}}$.  
    \item An interaction potential $V(\phi_{\mathbf{R}})$.  Typically, this potential includes a quadratic term, as well a cubic, quartic, etc.\ interactions among the scalars.  We denote the coefficient of the quadratic term as $m^{2}:$
    \begin{equation}
        V(\phi_{\mathbf{R}})=m^{2}|\phi_{\mathbf{R}}|^{2}+\cdots~.
    \end{equation}
\end{itemize}
 Given this data, our focus is the critical point defined by tuning the scalars to be massless $m^{2}\rightarrow0$.  In general, such a system is a gapless conformal field theory. It is also possible that the resulting system is first-order. 

 One way to understand the $G_{k}+ \phi_{\mathbf{R}}$ fixed point described above, is that it provides a transition between two topological phases.  These are achieved by activating the quadratic term in the potential and flowing to the infrared.  The result of these flows can be deduced by examining the initial gauge theory description of the fixed point:
\begin{itemize}
    \item $m^{2}>0:$ The scalar field $\phi_{\mathbf{R}}$ is massive and is integrated out.  At long distances, this results in the $G_{k}$ topological Chern-Simons theory.
    \item $m^{2}<0:$ The scalar field $\phi_{\mathbf{R}}$ condenses and Higgses the gauge group from $G$ to a subgroup $H$ stabilizing the expectation value.  At long distances, this results in a $H_{\tilde{k}}$ topological Chern-Simons theory.  Here, the level $\tilde{k}$ of $H$ is given by $\tilde{k}=nk$ where $n\in \mathbb{N}$ is the index of embedding of $H$ in $G.$\footnote{The index $n$ of the embedding $H \hookrightarrow G$ can be computed by taking any representation $R_G$ of $G,$ decomposing it into an $H$-representation $R_H$, and evaluating the ratio of Dynkin indices: $n = I_H/I_G$.}  
\end{itemize}
These two flows are shown below:
\begin{equation} \label{exampleflow}
    \begin{tikzcd}
	&& {G_{k}+ \phi_{\mathbf{R}} ~\text{Fixed Point}} \\
	{ H_{\tilde{k}}  } &&&& {G_{k}}
	\arrow[ from=1-3, to=2-1, "-|\phi_{\mathbf{R}}|^{2}" ', "\text{condensed}"]
	\arrow[from=1-3, to=2-5, "+|\phi_{\mathbf{R}}|^{2}" , "\text{massive}" ']
\end{tikzcd}.
\end{equation}
To make this discussion more precise, note that we should in fact view the potential deformations described above as relevant deformations of the Chern-Simons matter fixed point.  Thus, our discussion assumes that these deformations remain relevant at this fixed point and further that we can analyze their effects using the microscopic gauge theory Lagrangian. In other words, we assume throughout that the flow to the fixed point and the potential flows commute.   Additionally, since our analysis is concerned primarily with the nature of this fixed point, in the following we will describe energy scales relative to the flow diagram \eqref{exampleflow}.  Therefore, the ultraviolet will generally refer to the fixed point (not the weakly coupled gauge theory), and the infrared to the topological phases $G_{k}$ and $H_{\tilde{k}}.$  

\subsubsection{The Non-Relativistic Approximation}

The massive flow, $m^{2}>0,$ to $G_{k}$ provides a starting point to analyze the transition.  At long distances, the scalar field $\phi_{\mathbf{R}}$ may be viewed as an anyon $a_{\mathbf{R}}$ in this topological field theory: 
\begin{equation}
    \phi_{\mathbf{R}}\rightarrow a_{\mathbf{R}}~.
\end{equation}
More precisely, in the strict infrared, quanta of $\phi_{\mathbf{R}}$ have a parametrically large mass and therefore the number of them in any process is conserved.  Amplitudes with $N$ scalar particles may then be approximated in the TQFT as correlation functions involving $N$ anyon worldlines $a_{\mathbf{R}}$.  

We may reverse this idea to obtain a non-relativistic approximation to the massive flow valid at non-zero energies, which are small compared to the mass of $\phi_{\mathbf{R}}.$  Indeed, in this limit, the field $\phi_{\mathbf{R}}$ fluctuates and costs only finite (but large) energy to activate.  Thus, the vacuum should be viewed as containing a sum over virtual loops of scalar quanta, or equivalently loops of the anyon $a_{\mathbf{R}}$. It follows that we may approximate the partition function of the Chern-Simons matter fixed-point theory, $Z_{\text{CSM}},$  as a sum over insertions of loops of $a_{\mathbf{R}}.$ Schematically: 
\begin{multline}\label{nonrel}
    Z_{\text{CSM}}(m) \approx Z_{G_{k}}
    +\int \mathcal{D}L \; e^{-m\lvert L_i \rvert}Z_{G_{k}}(a_{\mathbf{R}}(L))\\
    +\int \mathcal{D}L_1 \mathcal{D}L_2 \; e^{-\sum_i m\lvert L_i\rvert}Z_{G_{k}}(a_{\mathbf{R}}(L_1),a_{\mathbf{R}}(L_2))+\cdots~,
\end{multline}
where $L_i$ is a worldline with length $\lvert L \rvert$ of the $i$-th anyon, and $\mathcal{D}L_i$ is an appropriate regularized path-integral measure over the space of loops.  In this approximation, each integrand  $Z_{G_{k}}(a_{\mathbf{R}}(L_1),\cdots)$ is a correlator in the topological phase $G_{k}$.

The non-relativistic approximation, can provide a useful first analysis of the fixed point.  For instance, it can be used to analyze scattering matrices and crossing symmetry of anyonic particles \cite{Jain:2014nza, Gabai:2022snc, Mehta:2022lgq}.

\subsection{Mesoscopic Models of Higgsing}\label{sec:lattice}

\begin{figure*}[t!]
    \centering
    \subfloat[]{\includegraphics[width=0.25\textwidth]{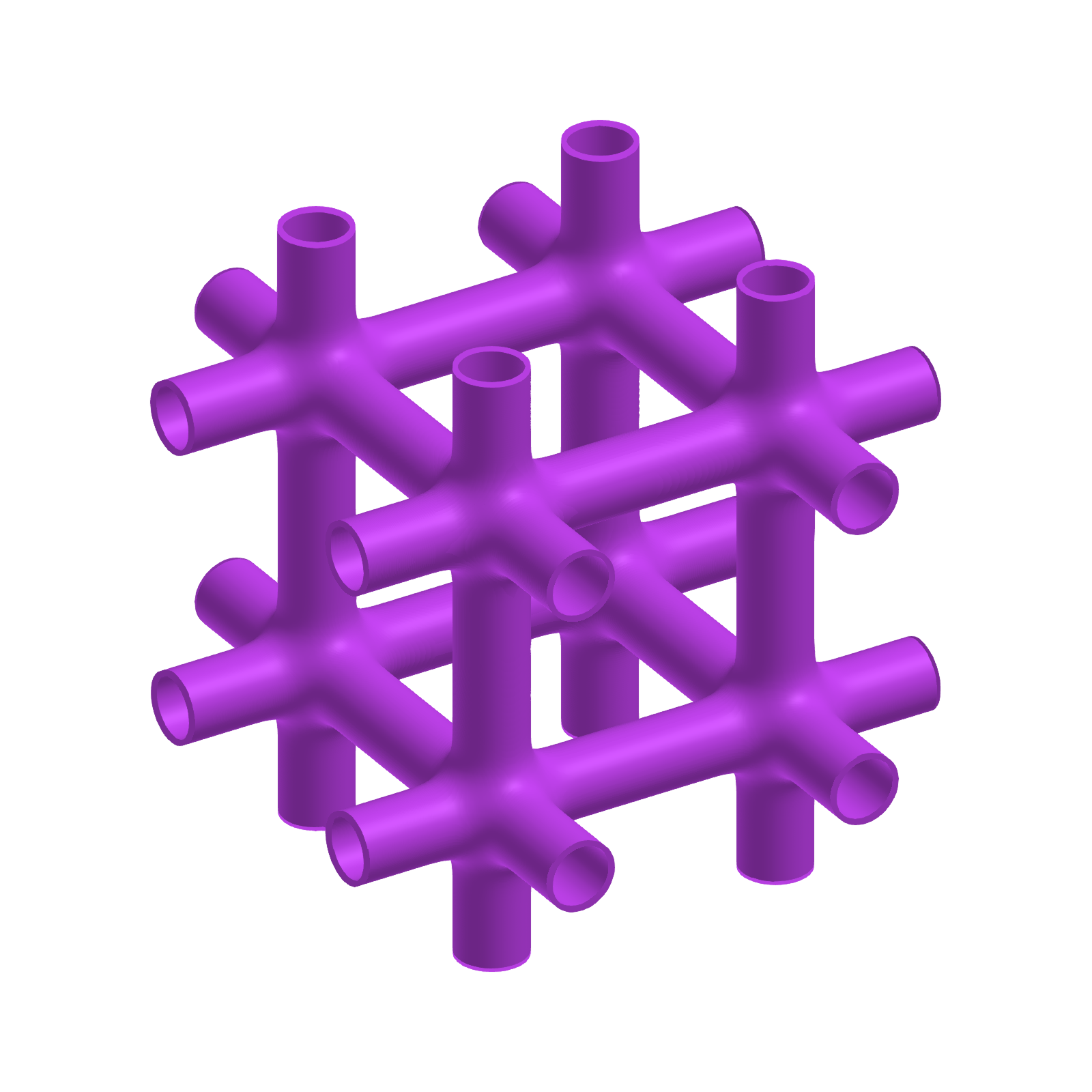}\label{fig:coset lattice a}}\hfill
    \subfloat[]{\includegraphics[width=0.25\textwidth]{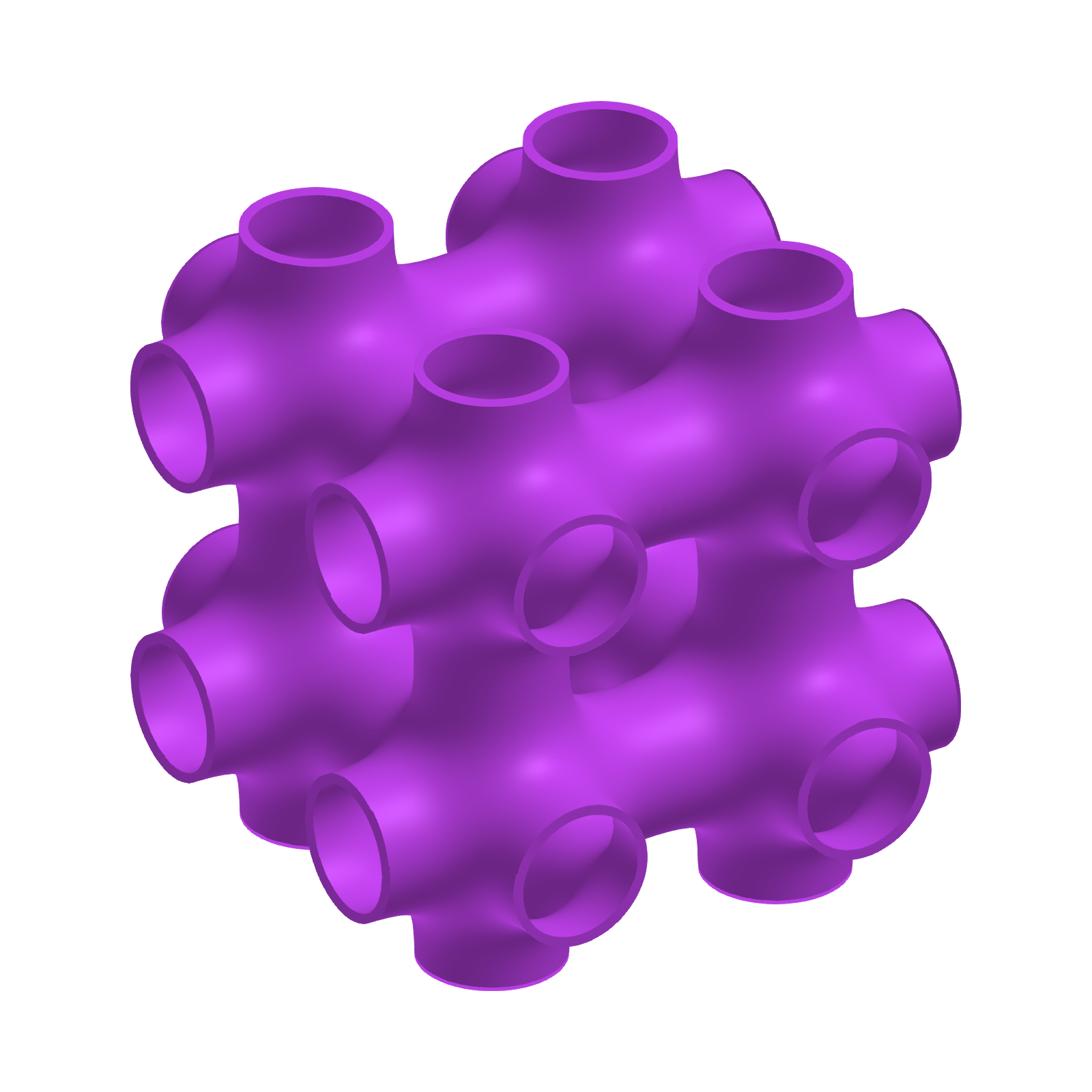}\label{fig:coset lattice b}}\hfill
    \subfloat[]{\includegraphics[width=0.25\textwidth]{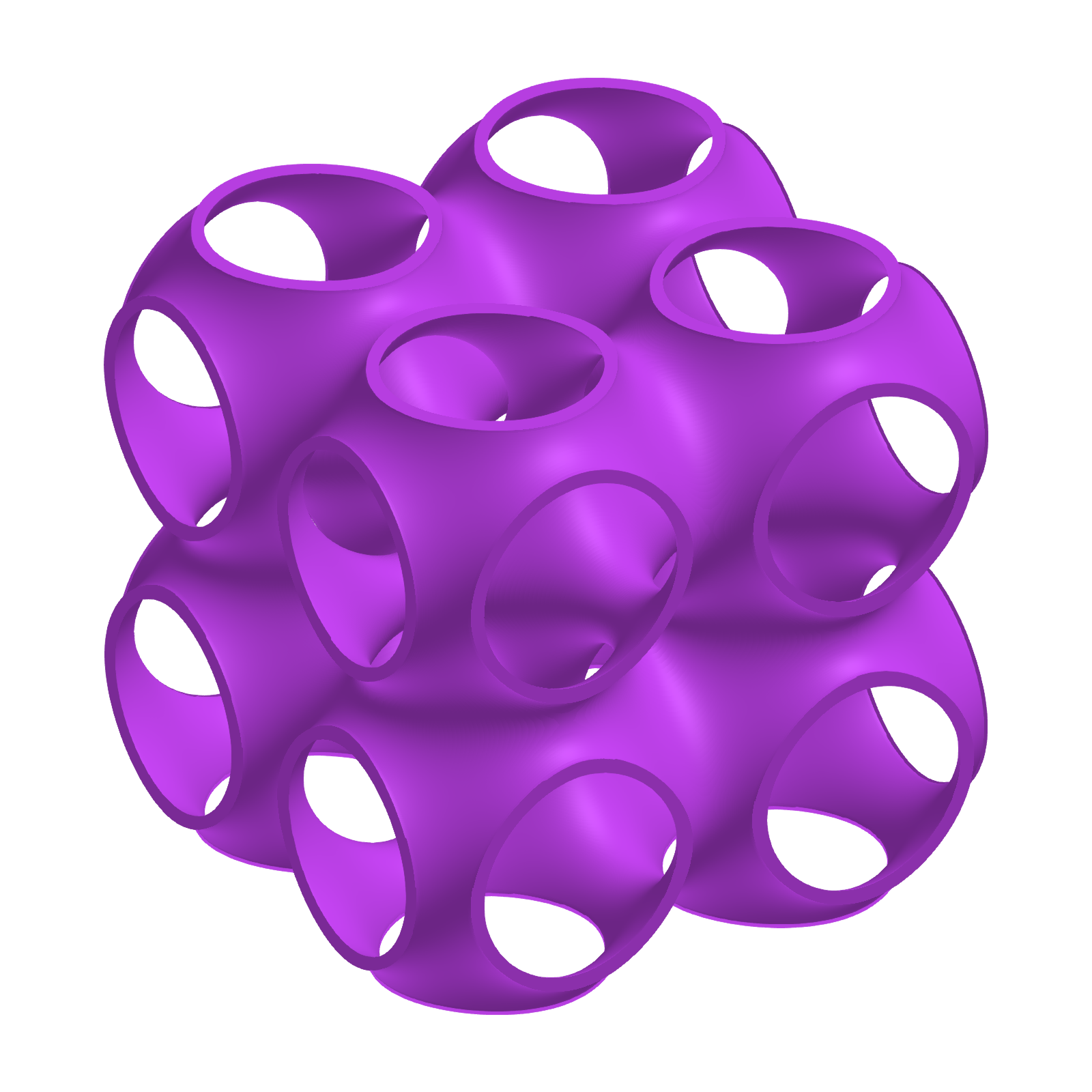}}
    \caption{A mesoscopic model of the Higgsing transition \eqref{exampleflow}.  In (a) we have intersecting cylindrical regions of radius $W$. The cylinders form a regular lattice as depicted, and the distance between two of the nearest parallel cylinders is $L$.  Outside the cylinders the theory is an a $G_{k}$ phase, inside it is in an $H_{\tilde{k}}$ phase.  At the interface is the coset $G_{k}/H_{\tilde{k}}$ which is in general gapless and chiral. See also Fig.~\ref{fig:coset lattice section} for the definitions of $L$ and $W$.  The width $W$ of the cylindrical regions is a parameter of the model.  For $W\rightarrow 0$ the theory is in a $G_{k}$ phase, while as $W\rightarrow L,$ (c), it is in an $H_{\tilde{k}}$ phase. }
    \label{fig:coset lattice 3d}
\end{figure*}

\begin{figure}[htbp]
    \centering
    \subfloat[]{\includegraphics[width=0.2\textwidth]{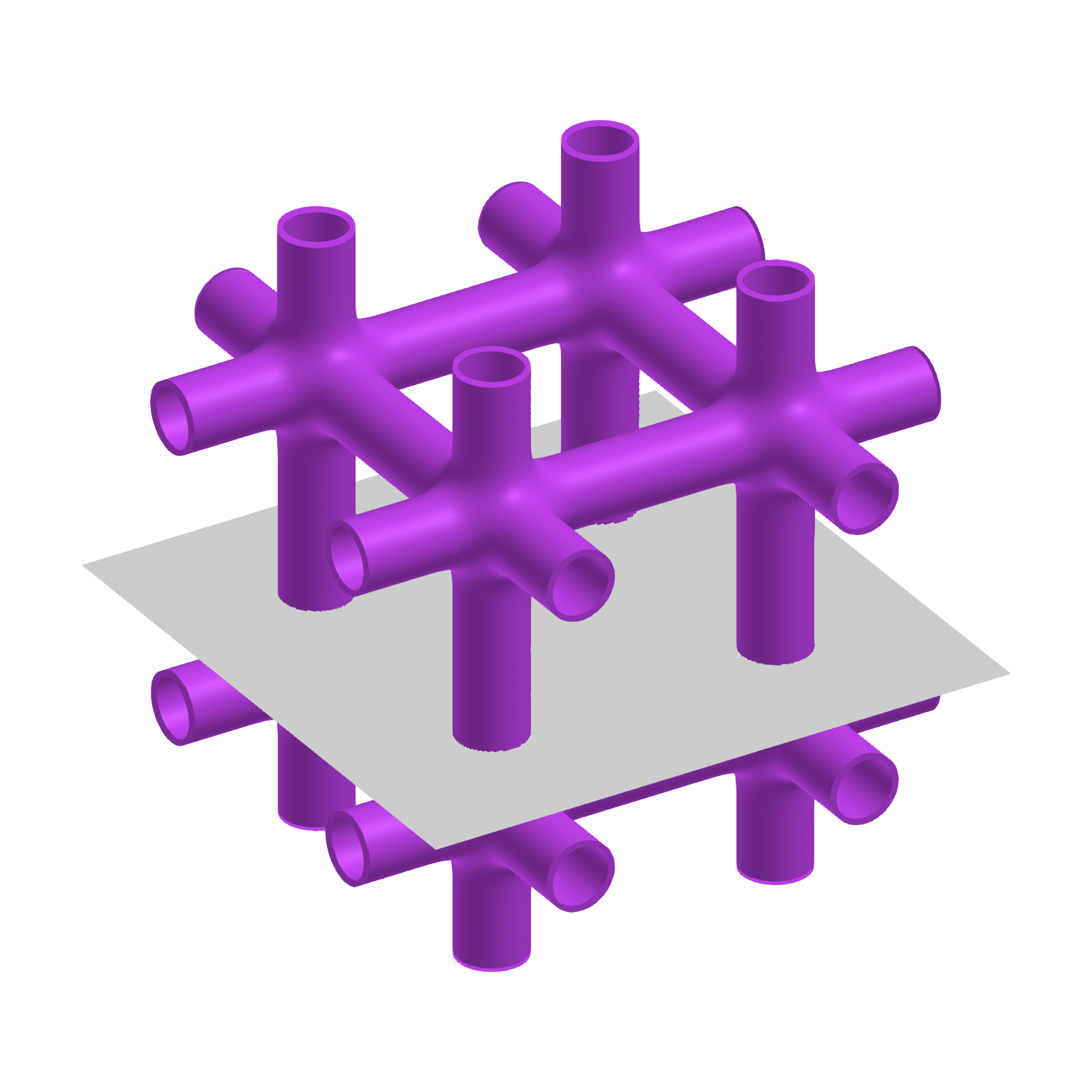}\hfill\label{fig:coset lattice section 3d}}
    \hfill
    \subfloat[]{\includegraphics[width=0.25\textwidth]{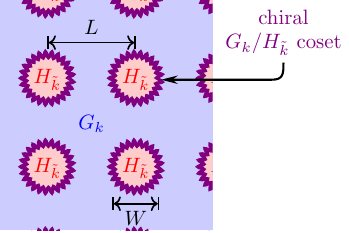}\label{fig:coset lattice section 2d}\hfill}
    \caption{A cross section (b) of the gray plane depicted in (a).}
    \label{fig:coset lattice section}
\end{figure}

To analyze the transition in more detail, we now construct a mesoscopic model that interpolates between the same two topological phases appearing in \eqref{exampleflow}. This model is Euclidean.  A related Hamiltonian model is presented below. To motivate the construction, we consider subjecting our Chern-Simons matter theory to a position (and time)-dependent mass term in the potential:
\begin{equation}
 m^2\rightarrow m^{2}(x)~.   
\end{equation}
  In the bulk of spacetime, $m^{2}(x)$ is assigned a large, positive value, whereas in cylindrical regions that intersect to form a cubical lattice, it takes a large, negative value. See Figure~\ref{fig:coset lattice a}.
  The interface between these two regions is a higher-genus Riemann surface, a domain wall network, where the mass parameter crosses zero.
  Upon flowing to the infrared, the region exterior to the cylinders is in the topological phase $G_{k}$, while the interior region is in the topological phase $H_{\tilde{k}}.$  See also Figure~\ref{fig:coset lattice section}.

Observe that the chiral central charges of $G_k$ and $H_{\tilde{k}}$ in general do not match.  Therefore, the domain wall interface between the bulk regions is in general chiral and thus gapless. 
We claim that this chiral mode is the (1+1)d coset theory $G_{k}/H_{\tilde{k}}$.  This model is the chiral sector of the gauged WZW model based on target space $G_{k}$ with gauge group $H_{\tilde{k}}.$ To see this, consider a Wilson line in the Chern-Simons matter theory in a $G$-representation, $\chi$, that crosses interface.
In the mesoscopic description, in the $H_{\tilde{k}}$ region the $G$ representation $\chi$ branches into a direct sum of $H$ representations.  Therefore, the domain wall modes should provide a junction (in general non-topological) between the $G_k$ Wilson line $\chi$ and the $H_{\tilde{k}}$ Wilson line $\psi$, as depicted in Fig.~\ref{fig: line and coset}(a) precisely when the branching of $\chi$ includes $\psi$.

The existence of these junctions is exactly the defining property of the chiral coset CFT.  Indeed, the representations (modules) of $G_k/H_{\tilde{k}}$, which we label by $\rho$, satisfy
\begin{equation} \label{branchingrules}
    \chi^{G_{k}}_{\chi}(q) = \sum_{\rho,\psi} n_{\chi}^{\rho \psi} \chi^{G_{k}/H_{\tilde{k}}}_{\rho}(q) \, \chi^{H_{\tilde{k}}}_{\psi}(q)~,
\end{equation}
with $q$ the modular parameter, and where $n_{\chi}^{\rho \psi} \in \mathbb{Z}^{+}$ dictates how the representations (modules) $\rho$ of the coset CFT couple to those of the $H_{\tilde{k}}$ WZW theory to yield the representation $\chi$ of $G_{k}$.\footnote{Notice that several copies of the trivial representation of the coset CFT $G_{k}/H_{\tilde{k}}$ with single~vacuum may appear in the branching decomposition. This signals the possible appearance of additional topological sectors.} This branching rule indicates that a vertex operator in the module $V_\rho$ can connect $\chi$ line in the $G_k$ phase and a $\psi$ line in the $H_{\tilde{k}}$ when $n_{\chi}^{\rho\psi}\neq0$. 

\subsubsection{A Model of the Quantum Field $\phi_{R}(x)$}

We can further motivate the mesoscopic model introduced above by comparing its features to those of the Chern-Simons matter fixed point.  

To begin, let us return to the non-relativistic expansion \eqref{nonrel} of the partition function.  In this description, it is clear that degrees of freedom are missing and have been integrated out along the renormalization group trajectory. Most directly we should ask: how can we produce the quantum field $\phi_{\mathbf{R}}(x)$? Of course, the particles created by $\phi_{\mathbf{R}}(x)$ are the anyons modeled by the line operator $a_{\mathbf{R}}$ in $G_{k}$.  In the IR these particles are infinitely massive and correspondingly, the lines are unbreakable.  In the fixed point however, the Wilson line in the representation $\mathbf{R}$ can end on the insertions of the field $\phi_{\mathbf{R}}(x).$  Note that this is compatible with the fact that $\phi_{\mathbf{R}}(x)$ is not gauge invariant: it is not a well-defined local operator but instead lives at the end of a line.  

\begin{figure}
    \includegraphics{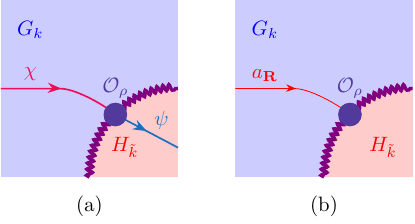}
    \caption{Lines transitioning between phases.  In (a) the transition between a general $G_{k}$ line $\chi$ to an $H_{\tilde{k}}$ line $\psi$ through a local operator $\mathcal{O}_{\rho}$ in the (1+1)d coset $G_{k}/H_{\tilde{k}}.$ In (b), a ray-shaped operator showing that in the mesoscopic model the anyon line $a_{\mathbf{R}}$ can end. These ray-shaped operators model the quantum field $\phi_{\mathbf{R}}.$  }
    \label{fig: line and coset}
\end{figure}

Now consider instead the mesoscopic model.  Here, in contrast, we will find a clear analog of the scalar field $\phi_{R}(x)$.  We start again with the Wilson line $a_{\mathbf{R}}$ in the same representation $\mathbf{R}$ as the scalar field. Note that the Higgsed group $H_{\tilde{k}}$ is characterized precisely by the fact that the branching of the $G$ representation $\mathbf{R}$ includes the trivial representation of $H$:
\begin{equation}
    \mathbf{R}|_{H}\cong \mathbf{1}_{H}+\cdots~.
\end{equation}
In terms of the coset character formula \eqref{branchingrules} this means that $n_{\mathbf{R}}^{\rho\mathbf{1}_{H}}\neq0$  for some module $\rho$ in the coset CFT. Thus, when the Wilson line $a_{\mathbf{R}}$ crosses the domain wall into the $H_{\tilde{k}}$ phase, it becomes transparent.  In other words, the line $a_{\mathbf{R}}$ can end on the interface as depicted in Fig.~\ref{fig: line and coset}(b), where it terminates on an operator $\mathcal{O}_\rho$ in the coset CFT. We regard these ray-shaped operators of $a_{\mathbf{R}}$ terminating on $\mathcal{O}_{\rho}$ as a proxy of the scalar field $\phi_{\mathbf{R}}(x)$.

Let us also inspect the two-point function of these operators in the limit $L\gg W$.  We consider a segment of the line $a_{\mathbf{R}}$ terminating at two points, $x_1$ and $x_2$, on the interface.  Then, the correlator contains a factor of the two-point function in the coset CFT on the interface surface $\Sigma$:
\begin{equation}\label{twopoint}
    \langle \mathcal{O}_\rho(x_1) \mathcal{O}_{\overline{\rho}}(x_2) \rangle_{\Sigma} \sim e^{-\frac{1}{W} h_\rho |x_1-x_2|}~,
\end{equation}
where $h_\rho$ is the conformal dimension of the primary $\mathcal{O}_\rho$.  Here, this estimate arises because the state on a circle created by $\mathcal{O}_{\rho}$ has energy $h_{\rho}/W$ and is propagated the distance $|x_1-x_2|$ before annihilating against the second insertion. On the other hand, we can compare \eqref{twopoint} to the expected exponential decay of the two-point correlator of the massive field $\phi_{R}(x)$. From this we can identify the scalar mass as:
\begin{equation}
    m\sim \frac{h_{\rho}}{W}~.
\end{equation}

Finally, we note that as we increase the width $W$ of the cylinders close to the lattice spacing $L$, it is no longer reliable to estimate the two-point function using radial quantization.  In particular, the exponential falloff in the coset two-point function will transition at short distances to power law behavior, indicating a potential gapless phase transition.

In summary, dialing $W$ from zero to $L$ in the mesoscopic model produces a phase transition between $G_{k}$ and $H_{\tilde{k}}.$ Our basic hypothesis is that this phase transition is in the same universality class as that of the underlying Chern-Simons matter fixed-point theory \eqref{exampleflow}. It is also possible that there are multiple intermediate phases as the parameters of the mesoscopic model are varied in which case we assume that one of the intermediate phases coincides with the fixed-point.

\subsubsection{Hamiltonian Model}

So far, we have described the mesoscopic model in a Euclidean framework with discrete time translation symmetry.  Here, we outline the corresponding Hamiltonian formulation with continuous time.  

The Hilbert space $\mathcal{H}$ is defined by the configuration in Figure~\ref{fig:coset lattice section 2d}.  In physical terms this consists of local coset modes from each cylinder-shaped ``lattice site" in the model. Additionally, we must take into account the fact that each cylinder appears at long distances to be a (non-simple) line inserted along time in the $G_{k}$ TQFT exterior.  Such lines induce additional states that we include. 

To write this more explicitly, let $r, s,\cdots$ be abstract indices labeling the lattice sites in the model. At each site the (1+1)d coset can be decomposed into modules $V_{\rho_{s}}$ (representations of the relevant chiral algebra).  We include in each site Hilbert space those summands which have a junction with the vacuum in $H_{\tilde{k}}$.  Physically, this corresponds to the fact that the spatial regions of $H_{\tilde{k}}$ are empty, i.e.\ they do not contain any non-trivial anyon inserted along time.  The full Hilbert space of the model is then a direct sum over all such sectors, indicated by  $\{\rho_{s}\},$ where each local factor is dressed by an appropriate line from the bulk $G_{k}$ Chern-Simons theory.  Explicitly:
\begin{equation}\label{hamiltonian}
    \mathcal{H} = \bigoplus_{\{\rho_s\}_s}\left( \mathrm{Hom}_{G_k}\left(\mathbf{1},\bigotimes_{r} \chi(\rho_r)\right)\otimes\bigotimes_{s} V_{\rho_s}\right)~,
\end{equation}
where $\mathrm{Hom}_{G_k}(\mathbf{1},\chi)$ is the Hilbert space of the $G_k$ Chern-Simons theory with an inserted timelike Wilson line $\chi$, and $\chi_{s}$ is:
\begin{equation}\label{chishrink}
\chi(\rho_s) = \sum_{\chi\in G_k} n_\chi^{\rho_s \mathbf{1}_{H}} \chi~.
\end{equation}
(The coefficients $n_\chi^{\rho\psi}$ are defined in \eqref{branchingrules}.) To understand \eqref{chishrink}, consider adiabatically shrinking the diameter of an $H_{\tilde{k}}$ cylinder while keeping it extended in the time direction. This results in the insertion of the line defined above at each site.\footnote{
The factor $\mathrm{Hom}_{G_k}\left(\mathbf{1},\otimes_s \chi(\rho_s)\right)$ in \eqref{hamiltonian} is for when the spatial manifold is $S^2$. For a general spatial manifold, the factor should be replaced by the result of quantizing the $G_k$ Chern-Simons theory with $\chi(\rho_s)$ inserted at each site and along time and hence depends on the global spatial topology.
}

The “free” part of the Hamiltonian is given by the sum of the coset conformal field theory Hamilitonians acting on each module $V_{\rho_s}$. Note that this implies that in the strict zero-size limit, the non-vacuum $\rho$ contributions in the Hilbert space \eqref{hamiltonian} acquire infinite energy $h_\rho/W\to\infty$ from the coset modes.  In particular, in this limit the model reduces to the expected $G_k$ Chern-Simons theory without additional insertions.

Beyond the free Hamiltonian, local interaction terms between neighboring sites mediated by anyon exchanges can be introduced. Schematically, such an interaction is represented as
\begin{equation}
    \includegraphics{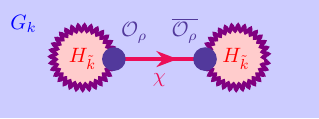}
    \label{eq: anyon exchange}
\end{equation}
where $\chi$ denotes an anyon in $G_k$ and $\mathcal{O}_{\rho}$ represents an operator in the module $V_\rho$ of the chiral coset theory, with $n_{\chi}^{\rho\mathbf{1}_{H}}\neq 0$. This can intuitively be understood as an infinitesimal version of one step time translation in the Euclidean model in Figure~\ref{fig:coset lattice a}. We expect that such interactions couple the cylinders effectively, thereby realizing the configurations illustrated in panels (b) and (c) of Figure~\ref{fig:coset lattice 3d} as the interaction strength is increased.

Here we note that the above construction cannot be used to realize the exact microscopic free Chern-Simons matter theory in any continuous limit. 
This is because, the model allows $G_k$ lines $\chi$ to have an end only when $n_{\chi}^{\rho\mathbf{1}_{H}}\neq 0$ for some $\rho$ in the coset CFT.
In the microscopic gauge theory, any $G_k$ line can end on a composite operator as long as it does not have a non-trivial electric one-form symmetry charge.
To realize that behavior, we further add the sectors including $H_{\tilde{k}}$ lines at the core of the cylinders. In the Hamiltonian picture, such new sector should be given an energy for each nontrivial line, modeling the massive excitations in the $H_{\tilde{k}}$ cylinder.
Our assumption is that such massive degrees of freedom are not important for the nature of phase transition between $G_k$ and $H_{\tilde{k}}$ phases.

We also emphasize that our model does \emph{not} provide a local way of constructing Chern-Simons theory Hilbert spaces; hence, it should not be interpreted as a microscopic model of the $G_k$ Chern-Simons theory.
Instead, we assume that the microscopic realization of the $G_k$ theory is given and use it as the basis for constructing a mesoscopic model of the transition to the $H_{\tilde{k}}$ phase.
In the extreme case where $G_k$ is trivial and the chiral coset theory is replaced by the corresponding $H_{\tilde{k}}$ WZW model, the total Hilbert space reduces to a direct product of identical infinite-dimensional spaces $V_{\mathbf{1}_{H}}$. This scenario corresponds to the setup of \cite{Sopenko:2023utk}, and we anticipate that an appropriate choice of interaction terms in \eqref{eq: anyon exchange} will yield a vacuum adiabatically connected to the state proposed in \cite{Sopenko:2023utk}.

\subsection{Anyon Condensation and Cosets}

The coset degrees of freedom on the domain walls provide us with a natural way to compare the lines in the two topological phases $G_{k}$ and $H_{\tilde{k}}$.  Indeed, for the examples that we will study below the (2+1)d TQFT $H_{\tilde{k}}$ admits the following presentation via anyon condensation:\footnote{\label{footinversion}More precisely, coset inversion (see \cite{Frohlich:2003hm, Cordova:2023jip}) says that, given a coset decomposition as in \eqref{branchingrules}, there exists an algebra $\mathcal{C}$ such that $H_{\tilde{k}}/\mathcal{C} = \left[G_{k}\times \left(\overline{\frac{G_{k}}{H_{\tilde{k}}}}\right)\right]/\mathcal{B}$. In our context, $\mathcal{C}$ will always be the trivial algebra.}
\begin{equation}\label{condense}
    H_{\tilde{k}}\cong \frac{\left[G_{k}\times \left(\overline{\frac{G_{k}}{H_{\tilde{k}}}}\right)\right]}{\mathcal{B}}~.
\end{equation}
Let us elaborate on the meaning of the above:
\begin{itemize}
    \item Each factor above is a (2+1)d TQFT.  In particular, $G_{k}/H_{\tilde{k}}$ is the coset TQFT which has as its one-sided chiral boundary the (1+1)d coset CFT.  The coset TQFT is in turn defined as follows:
    \begin{itemize}
        \item In simple cases the coset TQFT is a product Chern-Simons theory where the denominator has a time-reversed (negative) level yeilding:
        \begin{equation}\label{cosetnaive}
         \frac{G_{k}}{H_{\tilde{k}}} \equiv   G_{k}\times H_{-\tilde{k}}~.
        \end{equation}
        \item When the gauge groups $G$ and $H$ have a common center subgroup $Z$ we must modify the right-hand side of \eqref{cosetnaive} by quotienting by this common center factor.
        \item There may also be non-abelian bosons in the product Chern-Simons theory \eqref{cosetnaive} which must be condensed to accurately describe the TQFT. In complete generality, the principle defining the coset TQFT is that one should condense (gauge) the maximal braided fusion category which is common to both $G_{k}$ and $H_{\tilde{k}}.$  The case of the center quotient is an example of this principle when the common subfusion category is abelian.  
    \end{itemize}
    Examples of coset TQFTs which differ from \eqref{cosetnaive} by non-abelian anyon condensation are called \emph{maverick cosets} and play a prominent role below. See \cite{DiFrancesco:1997nk, Dunbar:1993hr, Pedrini:1999iy, Frohlich:2003hg,Frohlich:2003hm}.

    \item In \eqref{condense}, the barred factor has been time-reversed.  For the Chern-Simons factors, this flips the sign of the levels.  More generally, this conjugates the spins of all anyons. 
    
    \item In the denominator,  $\mathcal{B}$ is an algebra object of the TQFT. The notation means that certain anyons have been condensed i.e.\ gauged on the right-hand side.  Intuitively, this expression thus means that any line $\psi$ in the $H_{\tilde{k}}$ theory can be viewed as a pair $(\chi, \bar{\rho})$ in $G_{k}\times \overline{\frac{G_{k}}{H_{\tilde{k}}}}$ provided that we enforce the selection rules and identifications implied by non-abelian anyon condensation.  
\end{itemize}

In the above, the operation of non-abelian anyon condensation dictated by the object $\mathcal{B}$ is the least familiar. In brief this gauging operation is characterized by a finite formal sum of lines (non-simple anyon) \cite{Kong:2013aya} : 
\begin{equation}
    \mathcal{B}=\sum_{i}\beta_{i}~,
\end{equation}
where in our context, each $\beta_{i}$ appearing above is a pair:
\begin{equation}
    \beta_{i}=(\chi_{i}, \bar{\rho}_{i}) \in G_{k}\times \overline{\frac{G_{k}}{H_{\tilde{k}}}}~,
\end{equation}
and $\mathcal{B}$ includes the identity line with unit multiplicity in the sum.  Additionally, we require that $\mathcal{B}$ admit a fusion channel to itself dictated by a multiplication map $\mu$:
\begin{equation}
    \begin{tikzpicture}[baseline=1.2cm,scale=.6]
    \node[style=aR line, shape=coordinate] (U) at (0,3.7) {};
    \node[style=aR line, shape=coordinate] (J) at (0,2) {};
    \node[style=aR line, shape=coordinate] (L) at (-1,0) {};
    \node[style=aR line, shape=coordinate] (R) at (1,0) {};
    \draw[aR line, thick] (J) -- (L);
    \draw[aR line, thick] (J) -- (R);
    \draw[aR line, thick] (J) -- (U);
    \node[fill=black, circle, inner sep=0, minimum size=8pt] at (J) {};
    \node[anchor=south] at (U) {$\mathcal{B}$};
    \node[anchor=north east] at (L) {$\mathcal{B}$};
    \node[anchor=north west] at (R) {$\displaystyle\mathcal{B}=\oplus_i n_i \beta_i$};
    \node[anchor=west,xshift=5pt] at (J) {$\mu\in \mathrm{Hom}_{\mathcal{C}}(\mathcal{B}\otimes\mathcal{B},\mathcal{B})$};
\end{tikzpicture}
\scalebox{1.5}{:=}
\begin{tikzpicture}[baseline=1.2cm,scale=.6]
    \node[style=aR line, shape=coordinate] (U) at (0,3.7) {};
    \node[style=aR line, shape=coordinate] (J) at (0,2) {};
    \node[style=aR line, shape=coordinate] (L) at (-1,0) {};
    \node[style=aR line, shape=coordinate] (R) at (1,0) {};
    \draw[aR line, thick] (J) -- (L);
    \draw[aR line, thick] (J) -- (R);
    \draw[aR line, thick] (J) -- (U);
    \node[fill=black, circle, inner sep=0, minimum size=8pt] at (J) {};
    \node[anchor=south] at (U) {$ \beta_i$};
    \node[anchor=north east] at (L) {$ \beta_j$};
    \node[anchor=north west] at (R) {$\beta_k$};
    \node[anchor=west,xshift=5pt] at (J) {$\mu_{ij}^k$};
    \node at (-2,2) {\scalebox{1.3}{$\displaystyle\bigoplus_{i,j,k}$}};
\end{tikzpicture}
\end{equation}
Moreover, $\mathcal{B}$ must braid trivially with itself through the multiplication map:
\begin{equation}
    \begin{tikzpicture}[scale=.6,baseline=1.2cm]
    \node[style=aR line, shape=coordinate] (U) at (0,3.7) {};
    \node[style=aR line, shape=coordinate] (J) at (0,2) {};
    \node[style=aR line, shape=coordinate] (L) at (-1,0) {};
    \node[style=aR line, shape=coordinate] (R) at (1,0) {};
    \draw[aR line, thick] (J) -- (L);
    \draw[aR line, thick] (J) -- (R);
    \draw[aR line, thick] (J) -- (U);
    \node[fill=black, circle, inner sep=0, minimum size=8pt] at (J) {};
    \node[anchor=south] at (U) {$\mathcal{B}$};
    \node[anchor=north east] at (L) {$\mathcal{B}$};
    \node[anchor=north west] at (R) {$\displaystyle\mathcal{B}$};
    \node[anchor=west,xshift=3pt] at (J) {$\mu$};
\end{tikzpicture}
\scalebox{1.5}{=}
\begin{tikzpicture}[baseline=1.2cm,scale=.6]
    \node[style=aR line, shape=coordinate] (U) at (0,3.7) {};
    \node[style=aR line, shape=coordinate] (J) at (0,2) {};
    \node[style=aR line, shape=coordinate] (L) at (-1,0) {};
    \node[style=aR line, shape=coordinate] (R) at (1,0) {};
    \draw[aR line, thick] (J) .. controls (0.5,1) .. (L);
    \draw[draw=white, line width=6pt] (J) .. controls (-0.5,1) .. (R);
    \draw[aR line, thick] (J) .. controls (-0.5,1) .. (R);
    \draw[aR line, thick] (J) -- (U);
    \node[fill=black, circle, inner sep=0, minimum size=8pt] at (J) {};
    \node[anchor=south] at (U) {$\mathcal{B}$};
    \node[anchor=north east] at (L) {$\mathcal{B}$};
    \node[anchor=north west] at (R) {$\displaystyle\mathcal{B}$};
    \node[anchor=west,xshift=3pt] at (J) {$\mu$};
\end{tikzpicture}
\end{equation}
In particular, this implies that each line in the gauged algebra $\mathcal{B}$ must be a boson, i.e.\ it must have integral spin.\footnote{Beyond the conditions enumerated here, there are also algebraic identities that are required by the multiplication map $\mu$. We do not make use of these in the following, and refer to \cite{Fuchs:2002cm, Kong:2013aya, Kaidi:2021gbs, Cordova:2024goh} for details.}  

One consequence of this discussion is a formula for the total quantum dimension before and after gauging.  We recall that the total quantum dimension of a TQFT is sum over quantum dimensions squared of its simple lines, $\sum_{\ell \in \mathcal{I}} d_{\ell}^{2},$ where the individual quantum dimension of a simple line operator $\ell$ is defined as the expectation value of a trivial loop of $\ell$:
\begin{equation} \label{QuantumDimension}
    d_{\ell} = \adjustbox{valign=c}{\includegraphics[scale=0.8]{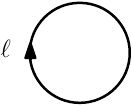}} ~.
\end{equation}
Defining the quantum dimension of an algebra as
\begin{equation}
    \mathrm{dim}(\mathcal{B}) = \sum_{i} d_{\beta_{i}}~,
\end{equation}
we then have the general formula:
\begin{equation}\label{dimform}
    \mathrm{dim}(H_{\tilde{k}}) = \frac{\mathrm{dim}(\left[G_{k} \times \left(\overline{\frac{G_{k}}{H_{\tilde{k}}}}\right)\right])}{\mathrm{dim}{(\mathcal{B})}^{2}}~.
\end{equation}
In practical examples below, we use these constraints to conjecture the existence of various algebras and apply them to analyze the behavior of topological lines across the Higgsing transition.

\subsection{Symmetries of the Higgsing Transition} \label{Section2D}

We now have the tools to analyze the symmetries of the Chern-Simons matter fixed point using the mesoscopic model framework.  We are particularly interested in (non-invertible) one-form symmetries.  Note that local operators are blind to such symmetries.  Therefore, all one-form symmetries must be visible in both gapped phases $G_{k}$ and $H_{\tilde{k}}$ which reside exterior and interior to the domain wall network in Figure \ref{fig:coset lattice 3d}.

To guide intuition, let us return again to the non-relativistic approximation \eqref{nonrel}. Consider any line/anyon $b$ in the theory $G_{k}.$  Since this theory is fully topological $b$ is of course also topological.  However, in general the line $b$ should be viewed as an emergent symmetry: it is topological in the strict IR, but not in the fixed-point theory where the scalar fluctuates.  A necessary condition for $b$ to remain topological at the fixed point can be deduced from \eqref{nonrel}.  Specifically $b$ must remain topological in the presence of the new vacuum state which contains a sum over insertions of $a_{\mathbf{R}}$.  For instance, this implies that loops of the line $a_{\mathbf{R}}$ must be transparent to loops of $b$.  Pictorially this means:
\begin{equation}\label{unlink}
    \mathord{
    \begin{tikzpicture}[scale=0.7, baseline=0pt]
      \draw[thick, red] (-0.5,1.2) -- (-0.5,0.7);
      \draw[thick, blue] (0.5,1.2) -- (0.5,0.7);
      \draw[thick, red] (-0.5,-0.5) -- (-0.5,-1.2);
      \draw[thick, blue] (0.5,-0.5) -- (0.5,-1.2);
      
      \draw[thick, blue] (0.5,0.7) .. controls (0.5,0.4) and (-0.5,0.4) .. (-0.5,0.1);
      \draw[white, line width=4pt] (-0.5,0.7) .. controls (-0.5,0.4) and (0.5,0.4) .. (0.5,0.1);
      \draw[thick, red] (-0.5,0.7) .. controls (-0.5,0.4) and (0.5,0.4) .. (0.5,0.1);
      
      \draw[thick, red] (0.5,0.1) .. controls (0.5,-0.2) and (-0.5,-0.2) .. (-0.5,-0.5);
      \draw[white, line width=4pt] (-0.5,0.1) .. controls (-0.5,-0.2) and (0.5,-0.2) .. (0.5,-0.5);
      \draw[thick, blue] (-0.5,0.1) .. controls (-0.5,-0.2) and (0.5,-0.2) .. (0.5,-0.5);
      
      \node[red] at (-0.8,1) {$a_{\mathbf{R}}$};
      \node[blue] at (0.8,1) {$b$};
    \end{tikzpicture}}
  =
    \mathord{
    \begin{tikzpicture}[scale=0.7, baseline=0pt]
      \draw[thick, red] (-0.5,1.2) -- (-0.5,-1.2);
      
      \draw[thick, blue] (0.5,1.2) -- (0.5,-1.2);
      
      \node[red] at (-0.8,1) {$a_{\mathbf{R}}$};
      \node[blue] at (0.8,1) {$b$};
    \end{tikzpicture}
    }
  \Longrightarrow \frac{S_{a_{\mathbf{R}},b}S_{0,0}}{S_{a_{\mathbf{R}},0}S_{b,0}}=1~,
\end{equation}
where above $S$ is the modular $S$-matrix of the $G_{k}$ topological theory. We note that the left-hand side above is a strictly stronger constraint involving the braiding $R$-matrix of the $G_{k}$ Chern-Simons theory.  We focus on its $S$-matrix implication (right-hand side) for simplicity of this intuitive discussion. In summary, equation \eqref{unlink} thus provides candidate lines $b$ that may, perhaps, remain topological in the  Chern-Simons matter fixed point.

\subsubsection{Local Modules and Topological Lines}

We can obtain more precision using the mesoscopic model of section \ref{sec:lattice} above.  The key point is that the anyon condensation procedure described above also gives us a complete picture of the lines that survive the gauging of an algebra $\mathcal{B}.$ Observe that after condensation, $\mathcal{B}$ becomes transparent and equivalent to the new identity line in the gauged theory ($H_{\tilde{k}}$ above in \eqref{condense}). Technically, the lines that characterize the theory after gauging are the so-called \textit{local modules} with respect to the algebra $\mathcal{B}$ in the original theory \cite{Kong:2013aya}. 

A general $\mathcal{B}$-module, $\mathcal{M},$ is defined diagrammatically as admitting a junction $\tilde{\mu}$ with the algebra $\mathcal{B}$:
\begin{equation}\label{module1}
    \begin{tikzpicture}[baseline=1.2cm,xscale=.6,yscale=.4]
    \node[style=aR line, shape=coordinate] (U) at (0,3.7) {};
    \node[style=aR line, shape=coordinate] (J) at (0,2) {};
    \node[style=aR line, shape=coordinate] (L) at (-1,0) {};
    \node[style=aR line, shape=coordinate] (R) at (0,0) {};
    \draw[aR line, thick] (J) -- (L);
    \draw[module line, thick] (J) -- (R);
    \draw[module line, thick] (J) -- (U);
    \node[fill=black, circle, inner sep=0, minimum size=8pt] at (J) {};
    \node[anchor=south] at (U) {$\mathcal{M}$};
    \node[anchor=north east] at (L) {$\mathcal{B}$};
    \node[anchor=north] at (R) {$\mathcal{M}$};
    \node[anchor=west,xshift=5pt] at (J) {$\tilde{\mu}\in \mathrm{Hom}_{\mathcal{C}}(\mathcal{B}\otimes\mathcal{M},\mathcal{M})$};
\end{tikzpicture}~.
\end{equation}
Mathematically, \eqref{module1} dictates how the algebra $\mathcal{B}$ can act on the module $\mathcal{M}$. Physically, $\mathcal{M}$ should be viewed as a candidate line after gauging $\mathcal{B}.$  Since $\mathcal{B}$ has become the identity in the gauged theory, it must admit a trivial fusion channel with $\mathcal{M}.$

Furthermore, for a $\mathcal{B}$-module $\mathcal{M}$ to be physically interpreted as a line in the theory after condensation, we require it to be \textit{local}. That is, to satisfy the condition
\begin{equation}\label{module2}
\begin{tikzpicture}[baseline=.8cm,xscale=.6,yscale=.4]
    \node[style=module line, shape=coordinate] (U) at (0,3.7) {};
    \node[style=module line, shape=coordinate] (J) at (0,2) {};
    \node[style=module line, shape=coordinate] (L) at (-1,0) {};
    \node[style=module line, shape=coordinate] (R) at (0,0) {};

    \coordinate (JR) at (0,1);
    \draw[module line, thick] (JR) -- (R);
    \draw[white, line width=5pt] (L) .. controls (.5,1-.2) .. (.5,1);
    \draw[aR line, thick] (L) .. controls (.5,1-.2) .. (.5,1) .. controls (.5,1+.2) and +(0,-.2) .. (-.5,1.5) .. controls +(0,.2) .. (J);
    \draw[white, line width=5pt] (J) -- (JR);
    \draw[module line, thick] (J) -- (JR);

    \draw[module line, thick] (J) -- (U);
    \node[fill=black, circle, inner sep=0, minimum size=8pt] at (J) {};
    \node[anchor=south] at (U) {$\mathcal{M}$};
    \node[anchor=north east] at (L) {$\mathcal{B}$};
    \node[anchor=north] at (R) {$\mathcal{M}$};
    \node[anchor=west,xshift=5pt] at (J) {$\tilde{\mu}$};
\end{tikzpicture}
\scalebox{1.5}{=}
\begin{tikzpicture}[baseline=.8cm,xscale=.6, yscale=.4]
    \node[style=module line, shape=coordinate] (U) at (0,3.7) {};
    \node[style=module line, shape=coordinate] (J) at (0,2) {};
    \node[style=module line, shape=coordinate] (L) at (-1,0) {};
    \node[style=module line, shape=coordinate] (R) at (0,0) {};

    \draw[module line, thick] (J) -- (R);
    \draw[aR line, thick] (J) -- (L);
    \draw[module line, thick] (J) -- (U);
    \node[fill=black, circle, inner sep=0, minimum size=8pt] at (J) {};
    \node[anchor=south] at (U) {$\mathcal{M}$};
    \node[anchor=north east] at (L) {$\mathcal{B}$};
    \node[anchor=north] at (R) {$\mathcal{M}$};
    \node[anchor=west,xshift=5pt] at (J) {$\tilde{\mu}$};
\end{tikzpicture}
\end{equation}
encapsulating the fact that a well-defined line in the theory after condensation must braid trivially with the algebra $\mathcal{B}$ since the latter is now transparent. 

Notice in particular the similarity of the diagrams in \eqref{unlink} and \eqref{module2}, where the putative symmetry $b$ is interpreted as the module $\mathcal{M},$ and $a_{\mathbf{R}}$ a part of the algebra $\mathcal{B}$ that is condensed.  This correspondence will be made precise below. 

\subsubsection{Extracting Topological Lines}\label{symalgorithm}

\begin{figure}
    \includegraphics{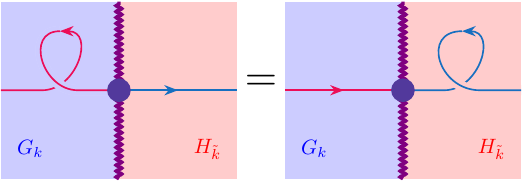}
    \caption{A topological line crossing the domain wall network.  If the junction is topological it has vanishing spin and hence the topological spins of the $G_{k}$ and $H_{\tilde{k}}$ lines match.}
    \label{fig: twist_propagate}
\end{figure}

Finally, we can now state a sharp proposal to use the mesoscopic model and anyon condensation to identify the lines which remain topological across a Higgsing transition.  We proceed as follows:
\begin{itemize}
    \item Identify a line $b \in G_{k}$ which braids trivially with the IR matter field line $a_{\mathbf{R}}\in G_{k}$ as in \eqref{unlink}.
    \item Present the Higgs phase of the Chern-Simons matter theory, i.e.\ $H_{\tilde{k}},$ via anyon condensation from the product $G_{k}\times \overline{\frac{G_{k}}{H_{\tilde{k}}}}$ as in \eqref{condense}.  Since $a_{\mathbf{R}}$ becomes dynamical, the condensing algebra $\mathcal{B}$ must contain as an element $(a_{\mathbf{R}}, \rho)$ for some $\rho$ in the coset TQFT.  This makes precise the discussion below \eqref{module2}: only by pairing $a_{\mathbf{R}}$ with coset degrees of freedom can it become bosonic and condense. 
    \item Embed the line $b$ into $G_{k}\times \overline{\frac{G_{k}}{H_{\tilde{k}}}}$ by assigning it trivial coset degrees of freedom.  In other words, identify $b\in G_{k}$ with $(b,\mathbf{1})\in G_{k}\times \overline{\frac{G_{k}}{H_{\tilde{k}}}}.$
    \item Check if $b$ remains a well-defined topological line after condensation so that it can be defined in the Higgsed phase $H_{\tilde{k}}.$ Specifically, we examine the fusion:
    \begin{equation} \label{fusionwithalgebra}
       \mathcal{B}\times (b, \mathbf{1})=(b,\mathbf{1}) + (a_{\mathbf{R}}\times b, \bar{\rho}) + \cdots~,
    \end{equation}
       where above we have extracted several lines known to appear above, but in general there are others in the ellipses. Observe:
       \begin{itemize}
           \item A necessary condition for $(b,\mathbf{1})$ to survive as a line in $H_{\tilde{k}}$ is that each simple line above that is not projected out by the module map $\tilde{\mu}$ of \eqref{module1} has the same topological spin.
           \item A sufficient condition for the above is that all elements on the right-hand side of \eqref{fusionwithalgebra} have the same topological spin, neglecting the module map projection.  In particular, when this is satisfied the details of the module map are immaterial.  (This is the case in our examples below.)
       \end{itemize}
 
\end{itemize}
The consistent spin condition discussed above has a simple graphical interpretation.  We consider the junction made by $b$ as it crosses the phase boundary from $G_{k}$ to $H_{\tilde{k}}.$ If the line $b$ is to be topological in both phases of the theory then the junction itself must be topological and hence in particular carry no spin.  See Figure~\ref{fig: twist_propagate}.

The result of the algorithm above are lines that are topological for all values of the parameters $L$ and $W$ in the mesoscopic model.  For instance, working in the Hamiltonian model we can see that any line $b$ identified above commutes with the interaction term \eqref{eq: anyon exchange}. This is because even if $b$ appears to be linked with the $\chi$ line in \eqref{eq: anyon exchange}, it can traverse the coset cylinder without altering the correlation functions, thereby becoming effectively unlinked. 

Assuming our basic hypothesis that the mesoscopic model transition is in the same universality class as the Higgsing transition, we are naturally led to conjecture that any such $b$ is topological throughout the Higgsing transition and in particular also at the Chern-Simons matter fixed point.  Below we will present a variety of examples of non-trivial one-form symmetries identified in this manner and show that such symmetries can in general be non-abelian.  

We emphasize that the argument presented in this section does not suggest that the microscopic $G_{k}$ Chern-Simons matter theory defined by the short-distance langrangian inherently realizes the one-form symmetry $b$. Instead, this symmetry is emergent, i.e.\ present only at the fixed point. Indeed, as remarked below \eqref{eq: anyon exchange} modeling the weakly-coupled gauge theory requires the inclusion of $H_{\tilde{k}}$ lines at the center of the
cylinders, which generally breaks the non-invertible one-form symmetry.

\subsubsection{Inverse Higgsing}

Before turning to a discussion of concrete examples, we note that our proposal in fact has a symmetry between the $H_{\tilde{k}}$ and $G_{k}$.  Indeed, consider a topological line in $\psi$ in $H_{\tilde{k}}.$ To test if it remains topological when we transition to the $G_{k}$ phase, we present the latter TQFT via anyon condensation as: 
\begin{equation}\label{inversion}
    G_{k}=\frac{H_{\tilde{k}}\times \frac{G_{k}}{H_{\tilde{k}}}}{\mathcal{A}}~,
\end{equation}
where as usual $\frac{G_{k}}{H_{\tilde{k}}}$ is the (2+1)d coset TQFT and $\mathcal{A}$ is a suitable algebra object inducing non-abelian anyon condensation.  We embed the line $\psi$ as $(\psi, \mathbf{1}) \in H_{\tilde{k}}\times \frac{G_{k}}{H_{\tilde{k}}}$, and then run the algorithm above to test whether $(\psi, \mathbf{1})$ remains topological across the phase boundary to $G_{k}.$  Happily in all our examples below we find perfect agreement between the symmetry algebras identified in this manner starting from either $G_{k}$ or $H_{\tilde{k}}.$

\subsection{Higgsing Transitions in Abelian Theories}

As a warmup example illustrating some of the ideas above let us consider the Higgsing transition in abelian gauge theory $U(1)_{k}$ driven by a complex scalar field of charge $q$, $\phi_{q}$.  The flow diagram is then given by:
\begin{equation}\label{abelianflow}
    \begin{tikzpicture}
        \node (UV) at (0,0) {$ {U(1)_{k}+ \phi_{q} ~\text{Fixed Point}}$};
        \node (TQFT1) at (-3.5,-1.5) {$(\mathbb{Z}_q)_k$};
        \node (TQFT2) at (3.5,-1.5) {$U(1)_k$};
        \draw[->] ($(UV.south west)!.5!(UV.south)$) -- node[midway,anchor=south east,font=\small] {$-|\phi_{q}|^{2}$} node[midway,anchor=north west,font=\small] {condensed} (TQFT1);
        \draw[->] ($(UV.south east)!.5!(UV.south)$) -- node[midway,anchor=south west,font=\small] {$+|\phi_{q}|^{2}$} node[midway,anchor=north east,font=\small] {massive} (TQFT2);
    \end{tikzpicture}
\end{equation}
Above, we note that the scalar condensate of charge $q$ Higgses the gauge group $U(1)$ to its $\mathbb{Z}_{q}$ subgroup that stabilizes the vacuum expectation value (since it does not act on $\phi_{q}$.)  Moreover, by the notation $(\mathbb{Z}_{q})_{k}$ we mean the discrete $\mathbb{Z}_{q}$ gauge theory with a Dijkgraaf-Witten term \cite{Dijkgraaf:1989pz, Kapustin:2014gua} interaction at level $k$. In practice this system can be efficiently realized as a $U(1)\times U(1)$ Chern-Simons theory with the level matrix, $M,$ and action:
\begin{equation}\label{twisteddiscrete}
    M = \begin{pmatrix}
k & q \\
q & 0
\end{pmatrix} \Longrightarrow S=\frac{iq}{2\pi}\int A \wedge dB +\frac{ik}{4\pi}\int A \wedge dA~.
\end{equation}
In the absence of the term proportional to $k$ above, \eqref{twisteddiscrete} is the standard presentation of a $\mathbb{Z}_{q}$ gauge theory. The level $k$ term provides the Dijkgraaf-Witten twist. We note that the theory $(\mathbb{Z}_{q})_{k}$ depends periodically on the level $k$ with period $2q:$
\begin{equation}\label{levelperiodic}
    (\mathbb{Z}_{q})_{\ell +2mq}\cong (\mathbb{Z}_{q})_{\ell}~.
\end{equation}
Below, we take $k$ to be even for simplicity which implies that the theory above is bosonic.  Finally, we also recall the spectrum of lines in $(\mathbb{Z}_{q})_{k}.$  These may be viewed as Wilson lines of the two gauge fields above with spin $h$ given by:
\begin{equation}\label{discretews}
    \exp\left(ix\oint A+iy\oint B\right)~, \hspace{.2in}h=\frac{xy}{q}- \frac{ky^{2}}{2q^{2}}~.
\end{equation}
Here $x,y$ are integers characterizing the charges of the line.  They are subject to identifications dictated by the level matrix:
\begin{equation}
    (x,y)\sim (x+q,y)~, \hspace{.2in} (x,y)\sim (x+k,y+q)~.
\end{equation}
Thus in general, there are $q^{2}$ lines.  Moreover, we can use the identifications above to determine that the fusion ring formed by the lines is:\footnote{This can be computed for instance, by noting that the Smith normal form of the level matrix $M$ in \eqref{twisteddiscrete} is:
\begin{equation}
    \mathrm{SNF}(M) = \begin{pmatrix}
\mathrm{gcd}(k,q) & 0 \\
0 & q^{2}/\mathrm{gcd}(k,q)
\end{pmatrix}~.
\end{equation}
}
\begin{equation}\label{fusionzq}
    \mathbb{Z}_{\mathrm{gcd}(k,q)}\times \mathbb{Z}_{q^{2}/\mathrm{gcd}(k,q)}~.
\end{equation}

\subsubsection{Transitions Without Symmetry: $\mathrm{gcd}(k,q) = 1$}

Take $k$ and $q$ coprime with $q<k$.  The Chern-Simons level $k$ in the $U(1)_{k}$ gauge theory truncates the abelian one-form symmetry to $\mathbb{Z}_{k}.$  The charged matter field $\phi_{q}$ is described by the anyon of charge $q,$ and hence is a generator of $\mathbb{Z}_{k}.$  In particular, all other non-identity lines braid non-trivially with this anyon.  Therefore in the semiclassical analysis of \eqref{unlink} there is no candidate symmetry line.  

Let us recover this conclusion from the Higgsed phase via coupling to the coset degrees of freedom.  We claim that in this case, the coset TQFT is simply:
\begin{equation}\label{trivialcoset}
   \frac{U(1)_{k}}{(\mathbb{Z}_{q})_{k}} \cong U(1)_{k}\times \overline{(\mathbb{Z}_{q})_{k}}\cong U(1)_{k}\times (\mathbb{Z}_{q})_{-k} ~.
\end{equation}
Note per the discussion below \eqref{condense},  we expect the coset TQFT to be a product of numerator and denominator (with opposite level) modulo an algebra of condensable common bosons.  Thus in \eqref{trivialcoset} we are asserting that this algebra of condensable common bosons is trivial.  To see this, observe from \eqref{fusionzq} that the total fusion ring in right-hand side of \eqref{trivialcoset} is
\begin{eqnarray}
    \mathbb{Z}_{k} \times \mathbb{Z}_{q^{2}}~,
\end{eqnarray}
and since $k$ and $q$ are coprime, there is no possible common subgroup in \eqref{trivialcoset} to condense.  Physically, since the coset TQFT \eqref{trivialcoset} is factorized, we expect its edge modes to also decouple into a chiral boson, the edge of $U(1)_{k}$, and a gapped sector, the edge of $(\mathbb{Z}_{q})_{-k}.$ 

Now we apply \eqref{condense} to present the Higgsed phase from anyon condensation in the massive phase:
\begin{eqnarray}\label{zqrecover}
    (\mathbb{Z}_{q})_{k} & \cong & \frac{U(1)_{k}\times \overline{\left(\frac{U(1)_{k}}{(\mathbb{Z}_{q})_{k}}\right)}}{\mathcal{B}}~, \nonumber \\
    & \cong & \frac{U(1)_{k}\times \left(U(1)_{-k}\times (\mathbb{Z}_{q})_{k}\right)}{\mathcal{B}}~,
\end{eqnarray}
where above, $\mathcal{B}$ is a suitable algebra.  However, since all the theories above are abelian $\mathcal{B}$ is simply the sum over elements of a subfusion ring each of which is bosonic.   Lines in the numerator of \eqref{zqrecover} can be labeled by quartets $(\ell_{1}, \ell_{2}, x, y)$ where $\ell_{i}=0, \cdots k-1$ indicates a charge in $U(1)_{k}$ and $(x,y)$ label the Wilson lines in $(\mathbb{Z}_{q})_{k}$ as in \eqref{discretews}. Adopting this notation it is straightforward to see that:
\begin{equation}
    \mathcal{B}=(0,0,0,0)+(1,1,0,0)+\cdots +(k-1,k-1,0,0)~.
\end{equation}
In particular, we see from this two expected features:
\begin{itemize}
    \item The condensation includes the anyon corresponding to the scalar field (the term $(q,q,0,0)$ above.) 
    \item There are no preserved topological lines in the Higgsed phase.  For instance we can apply the uniform spin criterion discussed in \eqref{fusionwithalgebra}.  A candidate line in $\ell \in U(1)_{k}$ is embedded in the numerator of the right-hand side of \eqref{zqrecover} as $(\ell,0,0,0)$.  Fusing with the algebra $\mathcal{B}$ gives:
    \begin{equation}
        \mathcal{B}\times (\ell,0,0,0)=\sum_{j=0}^{k-1}(\ell+j,j,0,0)~.
    \end{equation}
    We now compare the spin of the left-hand side to the spin of a simple anyon on the right-hand side.  Equality requires for all $j$:
    \begin{equation}
        \frac{\ell^{2}}{2k} = \frac{(\ell+j)^{2}}{2k}-\frac{j^{2}}{2k}~~\text{mod}~1~.
    \end{equation}
    This is clearly false so the symmetry is broken in the Higgsed phase as expected.
\end{itemize}

\subsubsection{Transitions With Symmetry: $k = nq$}

Next we consider a Higgsing transition where we expect preserved one-form symmetry.  We take $k=nq$ for $n$ a positive integer.  In this case the $\mathbb{Z}_{k}$ one-form symmetry of the Chern-Simons theory $U(1)_{k}$ should be broken down to the non-trivial subgroup $\mathbb{Z}_{q}\subseteq \mathbb{Z}_{k}.$

First, in the non-relativistic approximation of \eqref{unlink}, we recall that the braiding of lines in $U(1)_{k}$ with charges $\ell_{1}, \ell_{2}$ is:
\begin{equation}
    S_{\ell_{1},\ell_{2}}=\exp\left(\frac{2\pi i \ell_{1}\ell_{2}}{k}\right)~.
\end{equation}
Setting $\ell_{1}=q$, corresponding to the charged matter field, and $k=nq$, we see that the above is trivial precisely when $\ell_{2}$ is a multiple of $n$.  These lines generate the expected $\mathbb{Z}_{q}$ one-from symmetry group.

Now we recover this from the Higgsed phase.  In this case, we claim that the coset TQFT is:
\begin{equation}\label{cosetduality}
    \frac{U(1)_{nq}}{(\mathbb{Z}_{q})_{nq}} \cong U(1)_{nq}\cong U(1)_{k}~.
\end{equation}
In other words, the relevant edge modes are simply a chiral boson.  To derive this, we first note that in general we expect that the coset TQFT should arise from condensing the maximal common fusion subalgebra of the product Chern-Simons theory.  In this case, this results in:
\begin{equation}\label{cabcoset}
    \frac{U(1)_{nq}}{(\mathbb{Z}_{q})_{nq}}\cong \frac{U(1)_{nq}\times \overline{(\mathbb{Z}_{q})_{nq}}}{\mathcal{C}}~,
\end{equation}
where $\mathcal{C}$ is the abelian algebra corresponding the the $\mathbb{Z}_{q}$ common fusion algebra above.  Hence, equating \eqref{cabcoset} and \eqref{cosetduality} implies the relation:
\begin{equation}\label{cosetmagic}
    \frac{U(1)_{nq}\times \overline{(\mathbb{Z}_{q})_{nq}}}{\mathbb{Z}_{q}} \cong U(1)_{nq}~.
\end{equation}
This equivalence was derived in \cite{Cordova:2017vab} (See Appendix I).\footnote{This equivalence can be directly verified using the abelian anyon condensation techniques illustrated in the examples below.}  

Using the result \eqref{cosetmagic}, we now proceed to the Higgsed phase of our gauge theory.  From the general presentation \eqref{condense} we express:
\begin{eqnarray}\label{zqrecover2}
    (\mathbb{Z}_{q})_{qn} & \cong & \frac{U(1)_{qn}\times \overline{\left(\frac{U(1)_{qn}}{(\mathbb{Z}_{q})_{qn}}\right)}}{\mathcal{B}}~, \nonumber \\
    & \cong & \frac{U(1)_{qn}\times U(1)_{-qn}}{\mathcal{B}}~,
\end{eqnarray}
where the appropriate algebra $\mathcal{B}$ is the common $\mathbb{Z}_{n}$ subgroup above:
\begin{equation}
    \mathcal{B}=(0,0)+(q,q)+(2q,2q)+\cdots +((n-1)q,(n-1)q)~.
\end{equation}
Notice in particular the condensing algebra $\mathcal{B}$ contains the anyon corresponding to the scalar field in the term $(q,q)$ above.  

To verify \eqref{zqrecover2} we directly carry out the abelian anyon condensation.  The first step in this procedure is to determine which lines are not confined by the condensate $\mathcal{B}.$ These are all lines that braid trivially with the generator of $\mathcal{B}$ i.e.\ $(q,q)$.  The braiding of a general line $(\ell_{1},\ell_{2})$ with the condensing anyon $(q,q)$ is trivial if and only if:
\begin{equation}\label{constraint1}
    \frac{\ell_{1}q}{nq}-\frac{\ell_{2}q}{nq}=0 ~~\text{mod}~1\Longrightarrow \ell_{1}=\ell_{2}~~\text{mod}~n~.
\end{equation}
Thus, after condensation we must restrict the anyons obeying this condition.  Next, on this deconfined set, we must identify anyons which differ by fusion with the generator $(q,q)$.  Hence:
\begin{equation}\label{identify1}
    (\ell_{1},\ell_{2})\sim (\ell_{1}+q,\ell_{2}+q)~.
\end{equation}
The solution to the \eqref{constraint1} modulo the identification imposed by \eqref{identify1} leaves precisely $q^{2}$ lines which we may parameterize by equivalence classes represented by:
\begin{equation}\label{equivalenceclass}
    (r+sn,r)~, ~~~ r,s=0,\cdots, q-1~.
\end{equation}
This completes the condensation procedure.\footnote{\label{splitfoot}In general, anyon condensation requires a further step, where lines that are fixed under the fusion \eqref{identify1} split into distinct species in the theory after condensation \cite{Moore:1989yh, Hsin:2018vcg}. We do not encounter this here but it frequently occurs in more general examples.}  Notice that we have indeed found the correct number of lines to compare to $(\mathbb{Z}_{q})_{qn}$.  For instance by changing basis, we can verify the spins as follows.

     \emph{$n$ even}:  In this case the level periodicity \eqref{levelperiodic} implies $(\mathbb{Z}_{q})_{qn}\cong (\mathbb{Z}_{q})_{0}.$  We express the equivalence classes \eqref{equivalenceclass} in term of $x,y=0, 1, \cdots q-1$ as:
    \begin{equation}
        (r+sn,r)=\left(x+\frac{n}{2}y,x-\frac{n}{2}y\right)~,
    \end{equation}
    whose spin is:
    \begin{equation}
       h= \frac{\left(x+\frac{n}{2}y\right)^{2}}{2qn}-\frac{\left(x-\frac{n}{2}y\right)^{2}}{2qn}=\frac{xy}{q}~,
    \end{equation}
    exactly matching \eqref{discretews}.
    
     \emph{$n$ odd}:  Now, \eqref{levelperiodic} implies $(\mathbb{Z}_{q})_{qn}\cong (\mathbb{Z}_{q})_{q}.$ We express the equivalence classes \eqref{equivalenceclass} in term of $x,y=0, 1, \cdots q-1$ as:
    \begin{equation}
        (r+sn,r)=\left(x+\left(\frac{n-1}{2}\right)y,x-\left(\frac{n+1}{2}\right)y\right)~,
    \end{equation}
    whose spins are now:
    \begin{equation}
        h=\frac{\left(x+\left(\frac{n-1}{2}\right)y\right)^{2}}{2qn}-\frac{\left(x-\left(\frac{n+1}{2}\right)y\right)^{2}}{2qn}=\frac{xy}{q}-\frac{y^{2}}{2q}~,
    \end{equation}
    again matching \eqref{discretews}.

Finally, using the condensation presentation of the Higgsed phase in \eqref{zqrecover2}, we can see which lines remain topological across the Higgsing transition.  These are the lines of the form $(\ell,0)$ which survive the condensation procedure.  Matching with the equivalence classes \eqref{equivalenceclass} one sees that the charge $\ell$ is restricted to vanish modulo $n$ (i.e. the variable $r$ must be zero).  These anyons generate the expected $\mathbb{Z}_{q}$ preserved one-form symmetry.

\subsubsection{General $\gcd(k,q)$}

Let us briefly comment on the more general case when $\gcd(k,q) \neq 1.$  The analysis is similar to the example above and we omit derivations. 

In this case, the $\mathbb{Z}_{k}$ one-form symmetry of the pure Chern-Simons theory is screened down to $\mathbb{Z}_{\gcd(k,q)}$ by the dynamical scalar matter.  The relevant coset TQFT is now
\begin{equation}\label{gcdcoset}
    \frac{U(1)_{k}}{(\mathbb{Z}_{q})_{k}} \cong \frac{U(1)_{k} \times \overline{(\mathbb{Z}_{q})_{k}}}{\mathbb{Z}_{\mathrm{gcd}(k,q)}}~.
\end{equation}
Using the above, the presentation of the Higgsed phase via abelian anyon condensation as in \eqref{condense} is:
\begin{equation}
    (\mathbb{Z}_{q})_{k} \cong \bigg[ U(1)_{k} \times \overline{\bigg( \frac{U(1)_{k} \times \overline{(\mathbb{Z}_{q})_{k}}}{\mathbb{Z}_{\mathrm{gcd}(k,q)}}\bigg)} \bigg] / \mathbb{Z}_{\frac{k}{\mathrm{gcd}(k,q)} }~.
\end{equation}
In particular, the condensing algebra in the above includes the dynamical charge $q$ matter field in the $U(1)_{k}$ factor, i.e.\ $(q,\bar{\ell})$ for some line $\ell$ in the coset \eqref{gcdcoset}.  Following our algorithm described in Section \ref{Section2D} then reproduces the expected $\mathbb{Z}_{\gcd(k,q)}$ one-form symmetry.

For completeness, we also note that in this case the analog of \eqref{inversion} is
\begin{equation}
    U(1)_{k} \cong \bigg[ (\mathbb{Z}_{q})_{k} \times \bigg( \frac{U(1)_{k} \times \overline{(\mathbb{Z}_{q})_{k}}}{\mathbb{Z}_{\mathrm{gcd}(k,q)}}\bigg) \bigg] / \mathbb{Z}_{\frac{q^{2}}{\mathrm{gcd}(k,q)} }~,
\end{equation}
which can be used to reach the same conclusions about the one-form symmetry of this transition.

\subsection{Anomalies from Fusion Rules}
Here we collect some facts about non-invertible one-form symmetries and their anomalies. 
As mentioned previously, abstractly a finite one-form non-invertible symmetry in a (2+1)d system is described by topological line operators that form a braided fusion category $\mathcal{C}$. Specifically, it is a fusion category equipped with a braiding isomorphism $\sigma_{a,b}$ that maps $a \times b$ to $b \times a$ for each pair of objects $(a, b)$ in $\mathcal{C}$.

The Hopf link, composed of two simple lines $(a,b)$, defines the braiding $S$-matrix $S_{ab}$. When $S_{ab}$ is full-rank, the braided fusion category is called modular. When it is rank-one, or equivalently $\sigma_{ba}=\sigma_{ab}^{-1}$, the category is called symmetric.  In a Hopf link, the two lines do not intersect with each other and can remain far apart. Hence, the $S$-matrix of the topological lines must match along any renormalization group flow.  In particular this means that a non-trivial Hopf link of topological lines implies that the vacuum state has long-range entanglement. In other words, an $S$-matrix with a rank greater than one indicates a non-trivial anomaly, implying that the symmetry $\mathcal{C}$ cannot consistently act on the trivial theory.

When the symmetry is non-anomalous, i.e.\ the braiding is symmetric, the symmetry category $\mathcal{C}$ in a bosonic system must be equivalent to the representation category $\mathrm{Rep}(K)$ for some finite group $K$ \cite{Deligne:TensorCategories}.\footnote{For a fermionic theory, the theorem still holds with the group $K$ replaced by a ``supergroup'', which in this context means a finite group $K$ with an order two element identified as $(-1)^F$, the fermion parity.} See also \cite{nlab:deligne's_theorem_on_tensor_categories}. Conversely, if the one-form symmetry $\mathcal{C}$ has fusion rules that cannot be realized as $\mathrm{Rep}(K)$ for any finite group $K$, then we can conclude that the symmetry $\mathcal{C}$ is necessarily anomalous. This contrasts with the case of an abelian $\mathcal{C}$ (or an invertible symmetry), where the same fusion rule can admit both anomalous and non-anomalous braidings.

Relatedly, let us consider the case where the braided fusion category $\mathcal{C}$ contains a modular subcategory $\mathcal{D}$. Then the full category decomposes into a product \cite{Mueger:2003ModularCategories,Mueger:2003QuantumDouble,drinfeld2010braided}:
\begin{equation}
    \mathcal{C} \cong \mathcal{D}\times \mathcal{D}'~.
\end{equation}
(The case with abelian $\mathcal{D}$ and modular $\mathcal{C}$ was discussed in \cite{Hsin:2018vcg} from a physical perspective.) Here, $\mathcal{D}'$ is the M\"uger centralizer of $\mathcal{D}$ in $\mathcal{C}$. In particular, when the system is a TQFT acted on by the symmetry $\mathcal{C}$, a modular subcategory $\mathcal{D}$ indicates that the TQFT contains $\mathcal{D}$ as a decoupled factor. It is natural to expect that this decoupling continues to hold in a non-topological theory with a modular symmetry $\mathcal{D}.$

\section{Non-Abelian Higgsing \& Symmetry}\label{sec:ex}

We now turn to examples of Chern-Simons matter theories that conjecturally have non-invertible one-form symmetry.  We use the algorithm presented around \eqref{fusionwithalgebra} to check these proposals.\footnote{The spectrum and fusion rules of the MTCs used in this section can be obtained from the KAC software program \cite{KAC}.}

\subsection{Unitary Family: $PSU(2)_{-k}$ Symmetry}

Our first model is an $SU(k)$ gauge theory with matter in the symmetric rank two tensor representation.   
\begin{equation}\label{unituv}
    SU(k)_{2} + \phi_{\mbox{\tiny \yng(2)}} = SU(k)_{2} + \phi_{2 \mathbf{w}_{1}}~,
\end{equation}
where above, $\mathbf{w}_{1}$ indicates the weight labeling the fundamental representation of $SU(k).$ 

Our basic claim is that this model has topological symmetry lines $PSU(2)_{-k}$. Here, $PSU(2)_{k}$ means the subset of lines in the Chern-Simons theory $SU(2)_{k}$ which are neutral under the $\mathbb{Z}_{2}$ center symmetry.  More specifically, the $SU(2)_{k}$ fusion ring consists of $k+1$ simple lines labeled from $0$ to $k$, with fusion rules
\begin{equation}
    j_{1} \times j_{2} = \sum_{j = |j_{1}-j_{2}|}^{\mathrm{min}(j_{1}+j_{2},2k-j_{1}-j_{2})} j~, \label{SU2kFusionRules}
\end{equation}
where $j, j_{1},j_{2} = 0, 1, \ldots, k$, and the sum is restricted such that $j_{1} + j_{2} - j$ is even. The topological spins are given by $\theta_{j} = \exp(2 \pi i \frac{j(j+2)}{4(k+2)})$. The fusion ring $PSU(2)_{k}$ is defined as the closed subfusion ring corresponding to lines of even $j$ above. The preserved symmetry $PSU(2)_{-k}$ has spins which are complex conjugates of these.  
We note that for $k$ odd, $PSU(2)_{-k}$ is a well-defined fully modular-invariant TQFT on its own.  Meanwhile, for $k$ even, this symmetry is not modular. 

The Chern-Simons matter theory \eqref{unituv}, has a Higgsing transition given by the following flow diagram:
\begin{equation} \label{unitarymaverickflow}
    \begin{tikzcd}
	&& {SU(k)_{2} + \phi_{\mbox{\tiny \yng(2)}}} \\
	{ SO(k)_{4}  } &&&& {SU(k)_{2}}
	\arrow[hook', from=1-3, to=2-1]
	\arrow[from=1-3, to=2-5]
\end{tikzcd}.
\end{equation}
Here, the Higgsing pattern is achieved by $\phi_{\mbox{\tiny \yng(2)}}$ acquiring an expectation value which is a maximal rank diagonal matrix, and $SO(k)\subset SU(k)$ is the stabilizer subgroup.  

To investigate the symmetries, we will make use of the coset:
    \begin{equation}
    \frac{SU(k)_{2}}{SO(k)_{4}}~, \quad c = \frac{2(k-1)}{k+2}~,
\end{equation}
where above we have also indicated the chiral central charge $c$ of the (1+1)d edge modes.   This coset is a maverick theory.  As reviewed above, this means that when we want to describe the (1+1)d theory using a bulk Chern-Simons theory, we must gauge non-abelian anyons.  Let us denote these non-abelian anyons by $\mathcal{C}.$ Then, according to \cite{Cordova:2023jip} we have the following duality of Chern-Simons theories:
\begin{equation}
    \frac{SU(k)_{2}\times SO(k)_{-4}}{\mathcal{C}}\cong \frac{SU(2)_{k} \times U(1)_{-2k}}{\mathbb{Z}_{2}} \equiv \mathrm{PF}_{k}~.
\end{equation}
Here $\mathrm{PF}_{k}$ stands for the (2+1)d parafermion TQFT at level $k$ described by the Chern-Simons theories above.  Their edge modes are the (1+1)d parafermion coset CFT $SU(2)_{k}/U(1).$ 

When applying our model of the Higgsing transition, we will encounter the following relations between (2+1)d TQFTs (compare with \eqref{condense} and \eqref{inversion} above):
\begin{equation}\label{sokfrac}
    SO(k)_{4} \cong \frac{SU(k)_{2} \times \overline{\mathrm{PF}}_{k}}{\mathbb{Z}_{k}}~,
\end{equation}
as well as:
\begin{equation} \label{UnitaryFamilyExtension}
    SU(k)_{2} = \frac{SO(k)_{4} \times \mathrm{PF}_{k}}{\mathcal{A}}~,
\end{equation}
where $\mathcal{A}$ is a suitable non-abelian algebra specified below.

\subsubsection{$k = 3$ and the Three States Potts Model}

We begin with the simplest non-trivial case $k=3.$   The flow diagram is:
\begin{equation} \label{SU3LV2FLOW}
    \begin{tikzcd}
	&& {SU(3)_{2} + \phi_{\mathbf{6}}} \\
	{ SO(3)_{4}  } &&&& {SU(3)_{2} }
	\arrow[hook', from=1-3, to=2-1]
	\arrow[from=1-3, to=2-5]
    \end{tikzcd}.
\end{equation}
The data for the $SU(3)_{2}$ and $SO(3)_{4}$ Chern-Simons theories are presented in Table \ref{su3lv2table} and Table \ref{so3lv4table} respectively. The coset $SU(3)_{2}/SO(3)_{4}$ is known to be the Three State Potts Model (TSPM) \cite{Dunbar:1993hr}, which we present in Table \ref{TSPMtable}:
  \begin{equation}
    \frac{SU(3)_{2}\times SO(3)_{-4}}{\mathcal{C}}\cong \frac{SU(2)_{3} \times U(1)_{-6}}{\mathbb{Z}_{3}} \equiv \mathrm{PF}_{3} \cong \mathrm{TSPM}~.
\end{equation}
The explicit anyon condensation relations we must consider are:
\begin{equation} \label{k3coset}
    SO(3)_{4} \cong \frac{SU(3)_{2} \times \overline{\mathrm{TSPM}}}{\mathbb{Z}_{3}}~,
\end{equation}
and
\begin{equation} \label{k3extension}
    SU(3)_{2} = \frac{SO(3)_{4} \times \mathrm{TSPM}}{\mathcal{A}_{3}}~.
\end{equation}
We aim to argue that the fixed point theory has $PSU(2)_{-3}$ symmetry.  Note that this symmetry has a unique non-invertible line $W$ (corresponding to the $\mathbf{2}$ of $SU(2)$) with Fibonacci fusion rule:
\begin{equation}
    W\times W =\mathbf{1}+W~.
\end{equation}

\begin{table}[t]
\centering
\begin{tabular}{|p{2cm}|p{3cm}|p{3cm}|}
\hline 
\multicolumn{3}{|c|}{$SU(3)_{2}$} \\
\hline
Line label & Quantum Dimension & Conformal Weight \\
\hline
$\mathbf{1}$ & $d_{\mathbf{1}}=1$ & $h_{\mathbf{1}}=0$ \\
$\mathbf{3}$ & $d_{\mathbf{3}}=\frac{1+\sqrt{5}}{2}$ & $h_{\mathbf{3}}=4/15$ \\
$\mathbf{\bar{3}}$ & $d_{\mathbf{\bar{3}}}=\frac{1+\sqrt{5}}{2}$ & $h_{\mathbf{\bar{3}}}=4/15$ \\
$\mathbf{8}$ & $d_{\mathbf{8}}=\frac{1+\sqrt{5}}{2}$ & $h_{\mathbf{8}}=3/5$ \\
$\mathbf{6}$ & $d_{\mathbf{6}}=1 $ & $h_{\mathbf{6}}=2/3$ \\
$\mathbf{\bar{6}}$ & $d_{\mathbf{\bar{6}}}=1$ & $h_{\mathbf{\bar{6}}}=2/3$ \\
\hline
\end{tabular}
\caption{$SU(3)_{2}$ data. The fusion rules of $SU(3)_{2}$ are those of Fibonacci $\times$ $\mathbb{Z}_{3}$, with $\mathbf{8}$ the generator of Fibonacci, and the non-trivial elements of the $\mathbb{Z}_{3}$ factor are $\mathbf{6}$ and $\mathbf{\bar{6}}$.}  \label{su3lv2table}
\end{table}

\begin{table}[!b]
\centering
\begin{tabular}{|p{2cm}|p{3cm}|p{3cm}| }
\hline 
\multicolumn{3}{|c|}{$SO(3)_{4}$} \\
\hline
Line label & Quantum Dimension & Conformal Weight \\
\hline
0 & $d_{0}=1$ & $h_{0}=0$ \\
2  & $d_{2} = (3+\sqrt{5})/2$ & $h_{2}=1/5$ \\
$4_{1}$  & $d_{4_{1}} = (1 + \sqrt{5})/2$ & $h_{4_{1}}=3/5$ \\
$4_{2}$  & $d_{4_{2}} = (1 + \sqrt{5})/2$ & $h_{4_{2}}=3/5$ \\
\hline
\end{tabular}
\caption{$SO(3)_{4}$ data. The fusion rules of $SO(3)_{4}$ are those of two copies of Fibonacci, with the Fibonacci generators being $4_{1}$ and $4_{2}$, and $2 = 4_{1} \times 4_{2}$.}  \label{so3lv4table}
\end{table}

\begin{table}[t]
\centering
\begin{tabular}{|p{2cm}|p{3cm}|p{3cm}| }
\hline 
\multicolumn{3}{|c|}{Three-State Potts Model} \\
\hline
Line label & Quantum Dimension & Conformal Weight \\
\hline
$\mathbf{1}$ & $d_{\mathbf{1}}=1$ & $h_{\mathbf{1}}=0$ \\
$\sigma_{1}$ & $d_{\sigma_{1}}=\frac{1+\sqrt{5}}{2}$ & $h_{\sigma_{1}}=1/15$ \\
$\sigma_{2}$ & $d_{\sigma_{2}}=\frac{1+\sqrt{5}}{2}$ & $h_{\sigma_{2}}=1/15$ \\
$\varepsilon$ & $d_{\varepsilon}=\frac{1+\sqrt{5}}{2}$ & $h_{\varepsilon}=2/5$ \\
$Z_{1}$ & $d_{Z_{1}}=1 $ & $h_{Z_{1}}=2/3$ \\
$Z_{2}$ & $d_{Z_{2}}=1$ & $h_{Z_{2}}=2/3$ \\
\hline
\end{tabular}
\caption{Three-State Potts Model data. The fusion rules of the TSPM are those of Fibonacci $\times$ $\mathbb{Z}_{3}$, with $\varepsilon$ the generator of Fibonacci, and the non-trivial elements of the $\mathbb{Z}_{3}$ factor are $Z_{1}$ and $Z_{2}$.}  \label{TSPMtable}
\end{table}

As a first check to see the lines preserved along the flow, we employ the non-relativistic analysis and calculate monodromies in $SU(3)_{2}$ and it is straightforward to check that only $\mathbf{8}$ is preserved by $\mathbf{6}$.   Note also that $\mathbf{8}$ has Fibonacci fusion rules.  

Next we provide a more detailed check of this symmetry using cosets and anyon condensation.  The denominator of \eqref{k3coset} is abelian, generated by the anyon:
\begin{equation}
    \mathcal{B}_{3} = (\mathbf{1}, \mathbf{1}) + (\mathbf{6},\overline{Z_{1}}) + (\mathbf{\bar{6}}, \overline{Z_{2}})~,
\end{equation}
so we only have to extract the lines that have vanishing charge under the $\mathbb{Z}_{3}$ in $SU(3)_{2}$. This is precisely the $\mathbf{8}$ of $SU(3)_{2}$. Notice that as discussed in Section \ref{Section2D}, the condensation includes the anyon corresponding to the scalar field triggering the Higgsing transition. In this example this is the $\mathbf{6}$.

Next we analyze the condensation \eqref{k3extension}. In this case there are two non-trivial bosons $(4_{1},\varepsilon)$, $(4_{2},\varepsilon) \in SO(3)_{4} \times \mathrm{TSPM}$. To saturate quantum dimension formula \eqref{dimform} we must choose only one of them. The choice is immaterial, however, since $4_{1}$ and $4_{2}$ are symmetric in $SO(3)_{4}$. We choose:
\begin{equation}
    \mathcal{A}_{3} = (0, \mathbf{1}) + (4_{1},\varepsilon)~.
\end{equation}
We compute the fusions:
\begin{eqnarray}
     \mathcal{A}_{3} \times (0,\mathbf{1}) &=&  (0,\mathbf{1}) + (4_{1},\varepsilon)~, \nonumber \\
     \mathcal{A}_{3} \times (2,\mathbf{1}) &=& (2,\mathbf{1}) + (4_{2},\varepsilon) + (2,\varepsilon)~, \\
    \mathcal{A}_{3} \times (4_{1},\mathbf{1}) &=& (4_{1},\mathbf{1}) + (0,\varepsilon) + (4_{1},\varepsilon)~,\nonumber \\
     \mathcal{A}_{3} \times (4_{2},\mathbf{1}) &=& (4_{2}, \mathbf{1}) + (2,\varepsilon)~.\nonumber
\end{eqnarray}
     
We observe that the only non-trivial line which satisfies the uniform spin condition discussed in \eqref{fusionwithalgebra} is $4_{2}.$ This again reproduces the Fibonacci symmetry identified form the massive RG flow.

\subsubsection{General $k$}

We now consider the case of general $k$ in \eqref{unitarymaverickflow}.  The analysis is similar and our presentation is brief.  The rank 2 symmetric tensor is a generator of the $\mathbb{Z}_{k}$ center symmetry of $SU(k)_{2}$.\footnote{\label{bigfoot}To see this, notice that the center of $SU(N)_{k}$ is isomorphic to the corresponding $\mathbb{Z}_{N}$ outer automorphism of the affine Dynkin diagram shifting all fundamental weights by the adjacent one. Since the identity, labeled by the extended Dynkin labels $[0,0,0,\ldots,0,k]$, must be in the center, it follows that the elements of the center one-form symmetry are always labeled by extended Dynkin labels with a unique non-zero entry with value $k$: $[0,0,\cdots,0,k,0,\cdots,0]$.}  Therefore, in the non-relativistic approximation of \eqref{unlink}, the candidate lines which are preserved are precisely the lines in $SU(k)_{2}$ which are neutral under the full $\mathbb{Z}_{k}$ one-form symmetry.  Utilize the  level rank duality:
\begin{equation}
    SU(k)_{2} = \frac{SU(2k)_{1} \times SU(2)_{-k}}{\mathbb{Z}_{2}}~.
\end{equation}
Upon projecting to the subfusion ring above which is neutral under the $\mathbb{Z}_{k}$ center of the left-hand side, the right hand side simplifies to $PSU(2)_{-k},$ our claimed general symmetry.  Moreover, since the coset \eqref{sokfrac} involves abelian anyon condensation, it simply reproduces this condition.      

Now we check this result using the condensation presentation of $SU(k)_{2}$ as in \eqref{UnitaryFamilyExtension}.  For odd $k$ the full algebra in $SO(k)_{4} \times \mathrm{PF}_{k}$ is given by
\begin{equation} \label{algebraAk}
    \mathcal{A}_{k} = (0,0)  + \big( (4 \mathbf{w}_{\sigma})_{1}, (k-1,0) \big)+\sum_{i=1}^{(k-3)/2} \big( (2 \mathbf{w}_{i})_{1}, (2i,0)  \big) ~.
\end{equation}
In the above, $w_{\sigma}$ is the weight of the unique spinor representation, $w_{i}$ are the non-spinor weights and both are split into two lines (indicated by the subscript) in $SO(k)_{4},$  and $(2i,0) \in \mathrm{PF}_{k} = \frac{SU(2)_{k} \times U(1)_{-2k}}{\mathbb{Z}_{k}}$ stands for the line with Dynkin label $2i$ in $SU(2)_{k}$ and trivial charge in the $U(1)$ factor.

 The preserved symmetry algebra can now be obtained by finding those lines in $SO(k)_{4}$ whose decomposition under fusion with $\mathcal{A}_{k}$ contains only terms of uniform spin as discussed around \eqref{fusionwithalgebra}. One checks that these lines are precisely\footnote{Essentially, this generalizes the fact that for $k=3$ the line preserved corresponds to the partner of the line that we use in the Frobenius algebra $\mathcal{A}_{3}$ in the extension \eqref{UnitaryFamilyExtension}.}:
\begin{equation} \label{linespreservedkodd}
    \{0, \quad (4 \mathbf{w}_{\sigma})_{2}, \quad (2 \mathbf{w}_{i})_{2} \}, \quad i=1, \ldots, \frac{(k-3)}{2}~.
\end{equation}
Happily, these again form the fusion algebra $PSU(2)_{-k}.$

Similarly, for even $k,$ the algebra in $SO(k)_{4} \times \mathrm{PF}_{k}$ is given by
\begin{align}
    \mathcal{A}_{k} & = (0,0) + \sum_{i=1}^{\frac{k}{2} - 2} \big( (2 \mathbf{w}_{i})_{1}, (2i,0)  \big) \nonumber \\ &+ \big( (2 \mathbf{w}_{s} + 2 \mathbf{w}_{c})_{1}, (k-2,0) \big)   + \big( (4 \mathbf{w}_{s}), (k,0) \big)~,
\end{align}
where now $w_{s}$ and $w_{c}$ are the weights of the two spinor representations, and $(4 \mathbf{w}_{s}) \cong (4 \mathbf{w}_{c}).$ Again checking the uniform spin condition \eqref{fusionwithalgebra} we find that the preserved lines in $SO(k)_{4}$ are:
: 
\begin{equation} \label{linespreservedkeven}
    \{0, \quad (4 \mathbf{w}_{s}), \quad (2 \mathbf{w}_{s} + 2 \mathbf{w}_{c})_{2} , \quad (2 \mathbf{w}_{i})_{2} \}~,
\end{equation}
where $i=1, \ldots, \frac{k}{2}-2$.  These again define $PSU(2)_{-k}.$

\subsubsection*{Consistency Check from Conformal Embeddings}\label{sec:confembed}

Let us provide a general consistency check on the proposed symmetries.  Consider the conformal embedding\footnote{The branching rules for the conformal embedding $SU(n)_{m} \times SU(m)_{n} \hookrightarrow SU(nm)_{1}$ have been studied e.g. in \cite{Nakanishi:1990hj, Altschuler:1989nm, Walton:1988bs}.}:
\begin{equation}
    SU(k)_{2} \times SU(2)_{k} \hookrightarrow SU(2k)_{1}~.
\end{equation}
This implies the existence of an anyon condensation formula:
\begin{equation}\label{Dcond}
    \frac{SU(k)_{2}\times SU(2)_{k}}{\mathcal{D}_{k}}\cong SU(2k)_{1}~. 
\end{equation}
Here, the algebra object $\mathcal{D}_{k}$ can be understood as follows.  Inside $SU(k)_{2}$ is the previously identified fusion sub-algebra $PSU(2)_{-k}$.  Then, $\mathcal{D}_{k}$ is defined by diagonal subset of lines inside $PSU(2)_{-k}\times PSU(2)_{k}$ above.\footnote{Small $k$ examples of this condensation were presented in \cite{Cordova:2023jip}.}  Note that since the paired lines have opposite spin, they are condensable bosons.  Moreover, we can check the above using the quantum dimension formula \eqref{dimform}. (See Appendix \ref{AppendixA} for an explicit calculation).

Similarly, we have a conformal embedding\footnote{The branching rules for this conformal embedding have been studied e.g. in \cite{cmp/1104274518}.}:
\begin{equation}
    SO(k)_{4}\times SU(2)_{k}\hookrightarrow Sp(2k)_{1}~,
\end{equation}
implying the anyon condensation formula:
\begin{equation}\label{simplet}
    \frac{ SO(k)_{4}\times SU(2)_{k}}{\mathcal{D}_{k}}\cong Sp(2k)_{1}~.
\end{equation}
Here we have abused notation and written the algebra object above also as $\mathcal{D}_{k}.$  The reason is that the object is again composed of the diagonal anyons inside the $PSU(2)_{-k}\times PSU(2)_{k}$ sub fusion ring on the left-hand side above.  More specifically, for odd $k$:
\begin{equation}
    \mathcal{D}_{k} = (0,0) + \sum_{i=1}^{(k-3)/2} \big( (2 \mathbf{w}_{i})_{2}, 2i  \big) + \big( (4 \mathbf{w}_{\sigma})_{2}, k-1 \big)~,
\end{equation}
while for even $k:$
\begin{align}
    \mathcal{D}_{k} = (0,0) + \sum_{i=1}^{\frac{k}{2} - 2} \big( (2 \mathbf{w}_{i})_{2}, 2i  \big) & + \big( (2 \mathbf{w}_{s} + 2 \mathbf{w}_{c})_{2}, k-2 \big) \nonumber \\[0.1cm] &+ \big( (4 \mathbf{w}_{s}), k \big)~.
\end{align}
(Compare to  equations \eqref{UnitaryFamilyExtension}, and \eqref{algebraAk} through \eqref{linespreservedkeven}).

Armed with these results, we now take the entire flow diagram \eqref{unitarymaverickflow}, tensor it with $SU(2)_{k}$ and condense the algebra object $\mathcal{D}_{k}:$
\begin{equation}\label{unitarymaverickflowtensor}
    \begin{tikzpicture}
        \node (UV) at (0,0) {$\frac{SU(k)_{2}\times SU(2)_{k}}{\mathcal{D}_{k}} + \phi_{\mbox{\tiny \yng(2)}}$};
        \node (TQFT1) at (-2.5,-1.5) {$\frac{SO(k)_{4}\times SU(2)_{k}}{\mathcal{D}_{k}}$};
        \node (TQFT2) at (2.5,-1.5) {$\frac{SU(k)_{2}\times SU(2)_{k}}{\mathcal{D}_{k}}$};
        \draw[left hook->] ($(UV.south west)!.5!(UV.south)$) -- (TQFT1);
        \draw[->] ($(UV.south east)!.5!(UV.south)$) -- (TQFT2);
    \end{tikzpicture}.
\end{equation}
Finally we use the fact that across the duality \eqref{Dcond} the generators of the $\mathbb{Z}_{k}$ abelian one-form symmetry must match.  
Therefore, across the duality the symmetric rank two tensor of $SU(k)_{2}$ maps to the antisymmetric rank two tensor of $SU(2k)_{1}.$ (See footnote \ref{bigfoot}.)  Hence it is natural to conjecture that the flow generated by the symmetric rank two tensor of $SU(k)$ maps after gauging to the flow generated by the antisymmetric rank two tensor of $SU(2k).$  Assuming this, and simplifying \eqref{unitarymaverickflowtensor} using the condensation formulas \eqref{Dcond} and \eqref{simplet} gives the flow: 

\begin{equation}\label{unitarytosymplecticflow}
    \begin{tikzpicture}
        \node (UV) at (0,0) {${SU(2k)_{1} + \phi_{\mbox{\tiny \yng(1,1)}}}$};
        \node (TQFT1) at (-2.5,-1.5) {$Sp(2k)_{1}$};
        \node (TQFT2) at (2.5,-1.5) {$SU(2k)_{1}$};
        \draw[left hook->] ($(UV.south west)!.5!(UV.south)$) -- (TQFT1);
        \draw[->] ($(UV.south east)!.5!(UV.south)$) -- (TQFT2);
    \end{tikzpicture}.
\end{equation}
Strikingly, \eqref{unitarytosymplecticflow} is indeed a consistent Chern-Simons matter flow.  The symplectic Higgsing pattern is generated by $\phi_{\mbox{\tiny \yng(1,1)}}$ assuming an expectation value which is a maximal rank antisymmetric tensor (the invariant symbol of $Sp(2k)$).  

The fact that we generate consistent RG flows by gauging the conjectured non-invertible symmetry $PSU(2)_{-k}$ in our unitary flows \eqref{unitarymaverickflow} provides a strong consistency check on our results.

\subsubsection{Interpretation via Symmetry TQFT}\label{symTQFT1}

The $PSU(2)_{-k}$ symmetry admits a symmetry TQFT description \cite{Freed:2022qnc,Kaidi:2022cpf}: coupling the $SU(2)_{-k}$ (2+1)d TQFT to a (3+1)d gauge theory with a discrete $\mathbb{Z}_2$ 2-form gauge field $B$. The bulk (3+1)d TQFT has an exponentiated action
\begin{equation}\label{z4act1}
    \exp\left(\frac{k\pi i}{2}\int_{X} \mathfrak{P}(B)\right)~,
\end{equation}
where $X$ is a four-manifold and $\mathfrak{P}(B)$ is the Pontryagin square:
\begin{equation}
    \mathfrak{P}(B)\in H^{4}(X,\mathbb{Z}_{4})~.
\end{equation}
 The bulk-boundary coupling identifies the $\mathbb{Z}_2$ one-form symmetry line in the $SU(2)_{-k}$ TQFT with the boundary of the Wilson surface of $B$. We note that all $PSU(2)_{-k}$ lines are confined to the (2+1)d topological boundary, and the bulk theory is invertible when $k$ is odd.

Mathematically, the bulk theory is the Crane-Yetter TQFT \cite{Crane:1994ji} for the braided fusion category $PSU(2)_{-k}$, invertible if and only if the category is modular \cite{brochier2021invertible}.

\subsection{Spin Family: $PSpin(2N)_{2}$ Symmetry} \label{SpinMaverickFamily}

Our next class of models are $Spin(2N)$ gauge theories with matter in the symmetric traceless rank two tensor representation.   
\begin{equation}\label{spinuv}
    Spin(2N)_{2} + \phi_{\mbox{\tiny \yng(2)}} = Spin(2N)_{2} + \phi_{2 \mathbf{w}_{1}}~,
\end{equation}
where above, $\mathbf{w}_{1}$ indicates the weight labeling the vector representation of $Spin(2N).$ We take the scalars to be real so that the matter content is minimal.

We claim that this Chern-Simons matter fixed point preserves a $PSpin(2N)_{2}$ fusion algebra.  Here, by $PSpin(2N)_{2}$, we mean the sub-fusion algebra of $Spin(2N)_{2}$ which is uncharged under the center of $Spin(2N)$.  (Recall that for $N$ even, the center of $Spin(2N)$ is $\mathbb{Z}_{2}\times \mathbb{Z}_{2}$ while for $N$ odd, the center is $\mathbb{Z}_{4}.$)  For small $N$, $PSpin(2N)_{2}$ coincides, as a fusion ring but not as a braided fusion ring, with the representation ring of the Dihedral group $D_{N}$.\footnote{In our conventions, $D_{N}$ is the dihedral group with $2N$ elements.}  For instance:
\begin{equation}\label{dihedral}
    PSpin(2N)_{2}=\text{Rep}(D_{N})~,~~ N=3, 4, 5, 6, 7~,
\end{equation}
However, for general $N$, we are unaware of any elementary formula for this fusion ring.\footnote{The pattern listed in \eqref{dihedral} breaks e.g.\ at $N=9$ where as a fusion ring $PSpin(18)_{2}\cong \mathrm{Rep}(\mathbb{Z}_{3} \rtimes D_{3}).$}  

The Higgsing transition of this Chern-Simons matter theory is described by the following flow diagram:
\[\begin{tikzcd}
	& {Spin(2N)_{2} + \phi_{2 \mathbf{w}_{1}}} \\
	{ \frac{Spin(N)_{2}^{2}}{\mathbb{Z}_{2}}  } && {Spin(2N)_{2}}
	\arrow[hook', from=1-2, to=2-1]
	\arrow[from=1-2, to=2-3]
\end{tikzcd}.\]
Here, the condensed phase is achieved when $\phi_{\mbox{\tiny \yng(2)}}$ acquires an expectation value which is a maximal rank traceless symmetric tensor 
with equal eigenvalues on the first $N\times N$ block and the second $N\times N$ block, For instance:
\begin{equation}
    \phi_{\mbox{\tiny \yng(2)}}\sim \left(\delta_{ij}-\delta_{i+N,j+N}\right)~,~~~i,j=1,\cdots,N~,
\end{equation}
with $(Spin(N)_{2}\times Spin(N)_{2})/\mathbb{Z}_{2}$ the stabilizer of the above.    In particular, the $\mathbb{Z}_{2}$ quotient identifies the $\mathbb{Z}_{2}$ center subgroups of each $Spin(N)$ factor which measure the vector representation. 

To investigate the symmetries, we will make use of the following (1+1)d chiral coset model:
\begin{equation}
    \frac{Spin(2N)_{2}}{\left(Spin(N)_{2} \times Spin(N)_{2}\right)/\mathbb{Z}_{2}}~, \quad c = 1~,
\end{equation}
where we have also indicated the chiral central charge.  It has been observed \cite{Cordova:2023jip} that this coset is equivalent to a rational point on the orbifold branch of $c=1$ theories.  Specifically:
\begin{equation}
     \frac{Spin(2N)_{2}}{\left(Spin(N)_{2} \times Spin(N)_{2}\right)/\mathbb{Z}_{2}}\cong U(1)_{2N}^{\text{orb}}~,
\end{equation}
where above, $U(1)_{2N}^{\text{orb}}$ denotes a $\mathbb{Z}_{2}$ gauging of $U(1)_{2N}.$  The spectrum of this theory is summarized in Table \ref{Orbtable}.  
\begin{table}[t]
\centering
\begin{tabular}{|p{2.5cm}|p{2.5cm}|p{3cm}| }
\hline 
\multicolumn{3}{|c|}{Orbifold Model $U(1)_{2N}^{\text{orb}}$} \\
\hline
Line label & Q. Dimension & Conformal Weight \\
\hline
$\mathbf{1}$ & $d_{\mathbf{1}}=1$ & $h_{\mathbf{1}}=0$ \\
$\Theta$ & $d_{\Theta}=1$ & $h_{\Theta}=1$ \\
$\sigma_{1}, \sigma_{2}$ & $d_{\sigma}=\sqrt{N}$ & $h_{\sigma_{1}}=h_{\sigma_{2}}=1/16$ \\
$\tau_{1}, \tau_{2}$ & $d_{\tau}=\sqrt{N}$ & $h_{\tau_{1}}=h_{\tau_{2}}=9/16$ \\
$\phi_{1}, \phi_{2},\cdots,\phi_{N-1}$ & $d_{\phi_{i}}=2$ & $h_{\phi_{i}}=i^{2}/4N$ \\
$\phi^{1}_{N}, \phi^{2}_{N}$ & $d_{\phi^{1}_{N}}=d_{\phi^{2}_{N}}=1 $ & $h_{\phi^{1}_{N}}=h_{\phi^{2}_{N}}=N/4$ \\
\hline
\end{tabular}
\caption{Data for the orbifold theory $U(1)_{2N}^{\text{orb}}$.  In the notation of \cite{DiFrancesco:1997nk}, this is Chern-Simons theory with gauge group $O(2)^{0}_{2N,0}.$  The model has a total of $N+7$ simple lines.}  \label{Orbtable}
\end{table}
Crucial for our analysis below, the orbifold theory $U(1)_{2N}^{\text{orb}}$ has abelian anyons whose fusion mimics the center of $Spin(N)_{2}$.  Specifically, for $N$ odd the abelian fusion ring is $\mathbb{Z}_{4}$
\begin{equation}
    \phi^{(i)}_{N}\times \phi^{(i)}_{N}=\Theta~, ~~\phi^{(1)}_{N} \times \phi^{(2)}_{N} = \phi^{(2)}_{N} \times \phi^{(1)}_{N} = \mathbf{1}~,
\end{equation}
while instead for $N$ even the abelian fusion ring is $\mathbb{Z}_{2}\times \mathbb{Z}_{2}$
\begin{equation}
    \phi^{(i)}_{N}\times \phi^{(i)}_{N}=\mathbf{1}~, ~~\phi^{(1)}_{N} \times \phi^{(2)}_{N} = \phi^{(2)}_{N} \times \phi^{(1)}_{N} = \Theta~.
\end{equation}

When applying our model of the Higgsing transition, we will encounter the following relations between (2+1)d TQFTs.  The analog of 
\eqref{condense} is:
\begin{equation}
\begin{cases}\label{spincases}
    \frac{Spin(N)_{2} \times Spin(N)_{2}}{\mathbb{Z}_{2}} \cong \frac{Spin(2N)_{2} \times U(1)^{\mathrm{Orb}}_{-2N}}{\mathbb{Z}_{4}}~, & N~ \text{odd}~,\\
    \frac{Spin(N)_{2} \times Spin(N)_{2}}{\mathbb{Z}_{2}} \cong \frac{Spin(2N)_{2} \times U(1)^{\mathrm{Orb}}_{-2N}}{\mathbb{Z}_{2}\times \mathbb{Z}_{2}}~, & N~ \text{even}~.
\end{cases}
\end{equation}
Similarly, the analog of \eqref{inversion} is:
\begin{equation}
    Spin(2N)_{2} = \frac{\left( \frac{Spin(N)_{2} \times Spin(N)_{2}}{\mathbb{Z}_{2}} \times U(1)^{\mathrm{Orb}}_{2N} \right)}{\mathcal{A}_{N}}~,
\end{equation}
where $\mathcal{A}$ is a non-abelian algebra described below.

\subsubsection{Symmetry Analysis}

\begin{table}[t]
\centering
\begin{tabular}{|p{2.5cm}|p{3cm}| }
\hline 
\multicolumn{2}{|c|}{Abelian Anyons in $Spin(2N)_{2}$} \\
\hline
Line label &  Conformal Weight \\
\hline
$\mathbf{1}$ &  $h_{\mathbf{1}}=0$ \\
$\chi^{v}$ &  $h_{\chi^{v}}=1$ \\
$\chi^{s}, \chi^{c}$  & $h_{\chi^{s}}=h_{\chi^{c}}=N/4$ \\
\hline
\end{tabular}
\caption{Abelian anyons in $Spin(2N)_{2}.$  The anyons $\chi^{s}$ and $\chi^{c}$ measure charges of spinor representations, while $\chi^{v}$ measures the vector representation.  }  \label{spintable}
\end{table}

We proceed with the symmetry analysis.  First, in the non-relativistic approximation described around \eqref{unlink} we note that the symmetric tensor matter in our model is the generator of the $\mathbb{Z}_{2}$ one-form symmetry which measures the number of vector indices modulo two.  (See footnote \ref{bigfoot} for related discussion.)  Therefore in the notation of Table \ref{spintable}, the dynamical matter field flows at long distances to an abelian anyon:
\begin{equation}
     \phi_{\mbox{\tiny \yng(2)}} \longrightarrow \chi^{v}~.
\end{equation}
The candidate symmetry lines are thus those that are neutral under braiding with $\chi^{v}$ and hence form $Spin(2N)_{2}/\mathbb{Z}_{2}.$

To proceed further, we use the anyon condensation argument of section \ref{symalgorithm}. Specifically, from \eqref{spincases} to proceed inside a cylinder of the Higgsed phase, we must condense the whole center of $Spin(2N)_{2}$.  Thus inside $Spin(2N)_{2}\times U(1)_{-2N}^{\text{orb}}$ we consider the algebra object
\begin{equation}
    \mathcal{B}=(\mathbf{1},\mathbf{1})+(\chi^{s},\overline{\phi_{N}^{1}})+(\chi^{c},\overline{\phi_{N}^{2}})+(\chi^{v},\overline{\Theta})~.
\end{equation}
Note that all spins of the abelian anyons in $Spin(2N)_{2}$ have paired with anyons in the coset $U(1)_{2N}^{\text{orb}}$ to form condensable bosons.  For $N$ odd $\mathcal{B}$ is a $\mathbb{Z}_{4}$ algebra, while for $N$ even it is a $\mathbb{Z}_{2}\times \mathbb{Z}_{2}$ algebra.  In summary then condensing $\mathcal{B}$ leads to a fusion algebra of $PSpin(2N)_{2}$ preserved and faithfully acting in the Higgsed phase. 

Analogously, we can investigate the symmetries by transitioning from the Higgsed phase to the massive phase as in \eqref{inversion}.  For odd $N$, the full algebra object, $\mathcal{A}_{N}$ in $(Spin(N)_{2} \times Spin(N)_{2})/\mathbb{Z}_{2} \times U(1)^{\mathrm{Orb}}_{2N}$ is:
\begin{align}
    & \mathcal{A}_{N} = (0,\mathbf{1}) + \big( (0, 2\mathbf{w}_{1}), \Theta \big) \nonumber \\ & + \sum_{i}^{\frac{N-1}{2}-1} \big( (\mathbf{w}_{i}, \mathbf{w}_{i})_{1}, \phi_{2i} \big) + \big( (2\mathbf{w}_{\sigma}, 2\mathbf{w}_{\sigma})_{1}, \phi_{N-1} \big)~.
\end{align}
In the above, $\mathbf{w}_{\sigma}$ is the weight of the unique spinor representation, $\mathbf{w}_{i}$ are the non-spinor weights and the splitting into two lines is indicated by a subscript.  Further, we have the identification $(0, 2 \mathbf{w}_{1}) \cong (2 \mathbf{w}_{1}, 0)$ in $(Spin(N)_{2} \times Spin(N)_{2}) / \mathbb{Z}_{2}$ and the notation to label the primaries of $U(1)^{\mathrm{Orb}}_{2N}$ is summarized in Table \ref{Orbtable}.

The lines preserved now correspond to the subset of lines in $(Spin(N)_{2} \times Spin(N)_{2}) / \mathbb{Z}_{2}$ which obey the uniform spin condition \eqref{fusionwithalgebra} upon fusion with $\mathcal{A}_{N}$.  These are precisely:
\begin{equation}
    \{(0,0), \quad (0, 2 \mathbf{w}_{1}), \quad (2 \mathbf{w}_{\sigma}, 2 \mathbf{w}_{\sigma})_{2}, \quad (\mathbf{w}_{i}, \mathbf{w}_{i})_{2} \}~,
\end{equation}
where $i=1, \ldots, \frac{(N-3)}{2}$. Again we see that these lines form the algebra $PSpin(2N)_{2}.$

Analogously for $N$ even, the full algebra $\mathcal{A}_{N}$ in $\frac{Spin(N)_{2} \times Spin(N)_{2}}{\mathbb{Z}_{2}} \times U(1)^{\mathrm{Orb}}_{2N} $ is:
\begin{align}
    \mathcal{A}_{N} &= (0,\mathbf{1}) + \big( (0, 2\mathbf{w}_{1}), \Theta \big) \nonumber \\ &+ \big( (2\mathbf{w}_{s}, 2\mathbf{w}_{s}), \phi^{(1)}_{N} \big) + \big( (2\mathbf{w}_{s}, 2\mathbf{w}_{c}), \phi^{(2)}_{N} \big) \nonumber \\ & + \sum_{i = 1}^{\frac{N}{2}-2} \big( (\mathbf{w}_{i}, \mathbf{w}_{i})_{1}, \phi_{2i} \big) \nonumber \\ & + \big( (\mathbf{w}_{s} + \mathbf{w}_{c}, \mathbf{w}_{s} + \mathbf{w}_{c})_{1}, \phi_{N-2} \big)~,
\end{align}
where now $\mathbf{w}_{s}, \mathbf{w}_{c}$ are the two chiral spinor weights and we have the identifications $(0, 2 \mathbf{w}_{1}) \cong (2 \mathbf{w}_{1}, 0)$, $(2 \mathbf{w}_{s}, 2 \mathbf{w}_{s}) \cong (2 \mathbf{w}_{c}, 2 \mathbf{w}_{c})$, and $(2 \mathbf{w}_{s}, 2 \mathbf{w}_{c}) \cong (2 \mathbf{w}_{c}, 2 \mathbf{w}_{s})$ in $(Spin(N)_{2} \times Spin(N)_{2}) / \mathbb{Z}_{2}.$ The preserved lines are those in $(Spin(N)_{2} \times Spin(N)_{2}) / \mathbb{Z}_{2}$ obeying the uniform spin condition \eqref{fusionwithalgebra}:
\begin{align}
    \{(0,0), \quad (0, 2 \mathbf{w}_{1}), \quad (2 \mathbf{w}_{s}, 2 \mathbf{w}_{s}), \quad (2 \mathbf{w}_{s}, 2 \mathbf{w}_{c}), \nonumber \\  (\mathbf{w}_{s} + \mathbf{w}_{c}, \mathbf{w}_{s} + \mathbf{w}_{c})_{2}, \quad (\mathbf{w}_{i}, \mathbf{w}_{i})_{2} \}~,
\end{align}
where $i=1, \ldots, \frac{N}{2}-2$. These again form the fusion algebra $PSpin(2N)_{2}.$

\subsubsection{Interpretation via Symmetry TQFT}

The $PSpin(2N)_{2}$ symmetry admits a symmetry TQFT  description \cite{Freed:2022qnc,Kaidi:2022cpf} analogous to that of section \ref{symTQFT1}.  We couple $Spin(2N)_{2}$ (2+1)d TQFT to a bulk (3+1)d gauge theory based on abelian surfaces whose details depend on the parity of $N$.

For $N$ odd, the bulk theory is based on a  discrete $\mathbb{Z}_4$ 2-form gauge field $B$.  The action for this gauge field is determined by the spin of the abelian anyons corresponding to the center of $Spin(2N)_{2}.$ (See Table \ref{spintable}.)  Specifically, the exponentiated action is given by 
\begin{equation}\label{z4act}
    \exp\left(\frac{N\pi i}{2}\int_{X} \mathfrak{P}(B)\right)~,
\end{equation}
Where $X$ is a four-manifold and $\mathfrak{P}(B)$ is the Pontryagin square:
\begin{equation}
    \mathfrak{P}(B)\in H^{4}(X,\mathbb{Z}_{8})~.
\end{equation}
Similarly, for even $N,$ there are two $\mathbb{Z}_{2}$ 2-form gauge fields $B_{1}$ and $B_{2}$.  The exponentiated action is:
\begin{equation}\label{z22act}
    \exp\left(\frac{N\pi i}{4}\int_{X} \mathfrak{P}(B_{1})+\frac{N\pi i}{4}\int_{X} \mathfrak{P}(B_{2})\right)~,
\end{equation}
where now 
\begin{equation}
    \mathfrak{P}(B_{i})\in H^{4}(X,\mathbb{Z}_{4})~.
\end{equation}
The bulk-boundary coupling identifies the $\mathbb{Z}_2$ one-form symmetry lines in the $Spin(2N)_2$ TQFT with the boundary of the Wilson surfaces of the bulk 2-form gauge fields.  

We observe that the spin of the abelian anyons in $Spin(2N)_{2}$ is never minimal\footnote{For $\mathbb{Z}_{N}$ anyons with $N$ even, the minimal spin is $1/2N$.} and correspondingly the theories based on \eqref{z4act} and \eqref{z22act} are never invertible. Thus, the symmetry lines $PSpin(2N)_{2}$ cannot decouple since they are not modular.

\subsection{Exceptional CSM: Ising Symmetry}

\begin{table}[t]
\centering
\begin{tabular}{|p{2cm}|p{3cm}|p{3cm}| }
\hline 
\multicolumn{3}{|c|}{$E_{7,2}$} \\
\hline
Line label & Quantum Dimension & Conformal Weight \\
\hline
$0 \, (\mathbf{1})$   & $d_{0}= 1 $ & $h_{0}=0$ \\
$2 \mathbf{w}_{6}  \, (\mathbf{1463})$  & $d_{2 \mathbf{w}_{6}}= 1 $ & $h_{2 \mathbf{w}_{6}}= 3/2$ \\
$\mathbf{w}_{7} \, (\mathbf{912})$  & $d_{\mathbf{w}_{7}}=\sqrt{2}$ & $h_{\mathbf{w}_{7}}= 21/16$ \\
$\mathbf{w}_{6} \, (\mathbf{56})$  & $d_{\mathbf{w}_{6}}=(1 + \sqrt{5})/\sqrt{2}$ & $h_{\mathbf{w}_{6}}= 57/80$ \\
$\mathbf{w}_{5} \, (\mathbf{1539})$ & $d_{\mathbf{w}_{5}}=(1 + \sqrt{5})/2$ & $h_{\mathbf{w}_{5}} = 7/5$ \\
$\mathbf{w}_{1}\, (\mathbf{133})$  & $d_{\mathbf{w}_{1}}=(1 + \sqrt{5})/2$ & $h_{\mathbf{w}_{1}} = 9/10$ \\
\hline
\end{tabular}
\caption{$E_{7,2}$ data. The fusion ring of $E_{7,2}$ is that of $\mathrm{Fib} \times \mathrm{TY}(\mathbb{Z}_{2})$, where the Fibonacci element is given by $\mathbf{w}_{5}$ and those of $\mathrm{TY}(\mathbb{Z}_{2})$ by $2 \mathbf{w}_{6}$ and $\mathbf{w}_{7}$.}  \label{E72table}
\end{table}

As a final example, we consider a Chern-Simons matter model with an exceptional gauge group $E_{7}.$ We recall that the fundamental representation of $E_{7}$ is the $\mathbf{56}$ and for our matter we choose a scalar field in the antisymmetric rank two tensor representation $\mathbf{1539}$.  
\begin{equation}
    E_{7,2} + \phi_{\mathbf{1539}} = E_{7,2} + \phi_{\mathbf{w}_{5}}~.
\end{equation}
This theory has a Higgsing transition described by the following flow diagram:

\begin{equation}\label{e7flowd}
    \begin{tikzcd}
	& {E_{7,2} + \phi_{\mathbf{1539}}} \\
	{ \frac{SU(2)_{2} \times Spin(12)_{2}}{\mathbb{Z}_{2}}  } && {E_{7,2}}
\arrow[hook', from=1-2, to=2-1]
	\arrow[from=1-2, to=2-3]
\end{tikzcd}~.
\end{equation}
Our basic claim is that this model has Ising fusion category symmetry, where the spins match those in  $Spin(5)_{1}.$

To clarify the group theory of the Higgsed phase, note that the embedding of $(SU(2)_{2} \times Spin(12)_{2})\mathbb{Z}_{2}$ in $E_{7}$ is characterized by the branching rule:
\begin{equation}
    \mathbf{56} \longrightarrow (\mathbf{2}, \mathbf{12}) + (\mathbf{1}, \mathbf{32})~.
\end{equation}
Therefore, denoting by $A$ in index in the $\mathbf{56}$, there is a channel above where this decomposes to a product $A\rightarrow \alpha i$ where $\alpha$ is doublet index in $SU(2)$ and $i$ a vector index in $Spin(12).$ Thus the antisymmetric tensor $\mathbf{1539}$ can acquire an expectation value
\begin{equation}
    \phi_{AB}\sim \varepsilon_{\alpha\beta}~\delta_{ij}~,
\end{equation}
which is a singlet in $(SU(2)_{2} \times Spin(12)_{2})/\mathbb{Z}_{2}.$  From now on, we assume that the scalar potential is tuned to achieve this Higgsing.  The spectrum of the massive phase $E_{7,2}$ is summarized in Table \ref{E72table}, while the spectrum in the Higgsed phase $(SU(2)_{2} \times Spin(12)_{2})/\mathbb{Z}_{2}$ is summarized in Table \ref{SU2SPIN12GaugedTable}.

To investigate this flow we will analyze the coset
\begin{equation} \label{E7Coset}
    \frac{(E_{7})_{2}}{(SU(2)_{2} \times Spin(12)_{2})/\mathbb{Z}_{2}}~, \quad c = 8/10~,
\end{equation}
The result of the coset \eqref{E7Coset} is the Tetracritical Ising model \cite{Pedrini:1999iy}, 
\begin{equation} 
    \frac{(E_{7})_{2}}{(SU(2)_{2} \times Spin(12)_{2})/\mathbb{Z}_{2}} \cong \mathrm{TetraIsing}~,
\end{equation}
whose spectrum is summarized in Table \ref{Tetratable}. Our symmetry analysis is then based on the anyon condensation patterns (compare to \eqref{condense} and \eqref{inversion}):
\begin{equation} \label{E72betacoset}
    \frac{SU(2)_{2} \times Spin(12)_{2}}{\mathbb{Z}_{2}}= \frac{E_{7,2} \times \overline{\mathrm{TetraIsing}} }{\mathcal{B}}~,
\end{equation}
as well as:
\begin{equation} \label{E72alfacoset}
    E_{7,2} = \frac{ \frac{SU(2)_{2} \times Spin(12)_{2}}{\mathbb{Z}_{2}} \times \mathrm{TetraIsing}}{\mathcal{A}}~,
\end{equation}
where $\mathcal{A}$ and $\mathcal{B}$ are algebras specified below.

\begin{table}[t]
\centering
\begin{tabular}{|p{2cm}|p{3cm}|p{3cm}| }
\hline 
\multicolumn{3}{|c|}{Tetracritical Ising Model} \\
\hline
Line label & Quantum Dimension & Conformal Weight \\
\hline
(1,1) & $d_{(1,1)}=1$ & $h_{(1,1)}=0$ \\
(4,1) & $d_{(4,1)}=1$ & $h_{(4,1)}=3$ \\
(4,2) & $d_{(4,2)}=\sqrt{3}$ & $h_{(4,2)}=13/8$ \\
(4,4) & $d_{(4,4)}=\sqrt{3}$ & $h_{(4,4)}=1/8$ \\
(4,3) & $d_{(4,3)}=2 $ & $h_{(4,3)}=2/3$ \\
(3,1) & $d_{(3,1)}=\phi$ & $h_{(3,1)}=7/5$ \\
(2,1) & $d_{(2,1)}=\phi$ & $h_{(2,1)}=2/5$ \\
(2,2) & $d_{(2,2)}= \phi \sqrt{3}$ & $h_{(2,2)}=1/40$ \\
(3,2) & $d_{(3,2)}= \phi \sqrt{3}$ & $h_{(3,2)}=21/40$ \\
(3,3) & $d_{(3,3)}=2 \phi$ & $h_{(3,3)}=1/15$ \\
\hline
\end{tabular}
\caption{Tetracritical Ising Model data. The operator labels follow the Kac labels notation convention \cite{DiFrancesco:1997nk}.}  \label{Tetratable}
\end{table}

\begin{table}[!b]
\centering
\begin{tabular}{|p{2cm}|p{3cm}|p{3cm}| }
\hline 
\multicolumn{3}{|c|}{$(SU(2)_{2} \times Spin(12)_{2})/\mathbb{Z}_{2}$} \\
\hline
Line label & Quantum Dimension & Conformal Weight \\
\hline
$(0,0)$ & $d_{(0,0)}=1$ & $h_{(0,0)}=0$ \\
$(0,2\mathbf{w}_{1})$ & $d_{(0,2\mathbf{w}_{1})}=1$ & $h_{(0,2\mathbf{w}_{1})}=1$ \\
$(0,2\mathbf{w}_{s})$ & $d_{(0,2\mathbf{w}_{s})}=1$ & $h_{(0,2\mathbf{w}_{s})}=3/2$ \\
$(2,0)$ & $d_{(2,0)}=1$ & $h_{(2,0)}=1/2$ \\
$(0,\mathbf{w}_{s})$ & $d_{(0,\mathbf{w}_{s})}=\sqrt{6} $ & $h_{(0,\mathbf{w}_{s})}=11/16$ \\
$(0,\mathbf{w}_{1} + \mathbf{w}_{c})$ & $d_{(0,\mathbf{w}_{1} + \mathbf{w}_{c})}=\sqrt{6}$ & $h_{(0,\mathbf{w}_{1} + \mathbf{w}_{c})}=19/16$ \\
$(0,\mathbf{w}_{4})$ & $d_{(0,\mathbf{w}_{4})}=2$ & $h_{(0,\mathbf{w}_{4})}=4/3$ \\
$(0,\mathbf{w}_{2})$ & $d_{(0,\mathbf{w}_{2})}=2$ & $h_{(0,\mathbf{w}_{2})}=5/6$ \\
$(1,\mathbf{w}_{c})_{1}$ & $d_{(1,\mathbf{w}_{c})_{1}}=\sqrt{3}$ & $h_{(1,\mathbf{w}_{c})_{1}}=7/8$ \\
$(1,\mathbf{w}_{c})_{2}$ & $d_{(1,\mathbf{w}_{c})_{2}}=\sqrt{3}$ & $h_{(1,\mathbf{w}_{c})_{2}}=7/8$ \\
$(1,\mathbf{w}_{1} + \mathbf{w}_{s})_{1}$ & $d_{(1,\mathbf{w}_{1} + \mathbf{w}_{s})_{1}}=\sqrt{3}$ & $h_{(1,\mathbf{w}_{1} + \mathbf{w}_{s})_{1}}=11/8$ \\
$(1,\mathbf{w}_{1} + \mathbf{w}_{s})_{2}$ & $d_{(1,\mathbf{w}_{1} + \mathbf{w}_{s})_{2}}=\sqrt{3}$ & $h_{(1,\mathbf{w}_{1} + \mathbf{w}_{s})_{2}}=11/8$ \\
$(1,\mathbf{w}_{1})$ & $d_{(1,\mathbf{w}_{1})}=2\sqrt{2}$ & $h_{(1,\mathbf{w}_{1})}=31/48$ \\
$(1,\mathbf{w}_{3})_{1}$ & $d_{(1,\mathbf{w}_{3})_{1}}=\sqrt{2}$ & $h_{(1,\mathbf{w}_{3})_{1}}=21/16$ \\
$(1,\mathbf{w}_{3})_{2}$ & $d_{(1,\mathbf{w}_{3})_{2}}=\sqrt{2}$ & $h_{(1,\mathbf{w}_{3})_{2}}=21/16$ \\
\hline
\end{tabular}
\caption{$(SU(2)_{2} \times Spin(12)_{2})/\mathbb{Z}_{2}$ data.}  \label{SU2SPIN12GaugedTable}
\end{table}

\subsubsection{Symmetry Analysis}

We begin with our analysis of the preserved topological lines using the condensation formula \eqref{condense}.  By analyzing the quantum dimension formula \eqref{dimform} it is easy to see that the only interesting candidate for $\mathcal{B}$ is\footnote{The algebra $\mathcal{B}' = (0, \overline{(1,1)}) + (\mathbf{w}_{5}, \overline{(2,1)})$ also saturates quantum dimension, but it is related by the $\mathbb{Z}_{2}$ symmetry of the Tricritical Ising Model to $\mathcal{B}$, and leads to no new conclusions.}
\begin{equation}
    \mathcal{B} = (0, \overline{(1,1)}) + (\mathbf{w}_{5}, \overline{(3,1)})~.
\end{equation}
The line $\mathbf{w}_{5}=\mathbf{1539}$ in $E_{7,2}$ (the IR limit of the matter field) is a Fibonacci line. Notice that this line corresponds to the anyon in the same representation as that of the scalar field triggering the Higgsing transition, as expected from the discussion in Section \ref{Section2D}.

We can see which lines in $E_{7,2}$ remain topological using the uniform spin condition \eqref{fusionwithalgebra} leading to the topological spectrum $0,2 \mathbf{w}_{6}, \mathbf{w}_{7}.$  Indeed, from:
\begin{eqnarray}
    \mathcal{B} \times (2 \mathbf{w}_{6}, \overline{(1,1)}) & = & (2 \mathbf{w}_{6}, \overline{(1,1)}) + (\mathbf{w}_{1}, \overline{(3,1)})~,\\
      \mathcal{B} \times (\mathbf{w}_{7}, \overline{(1,1)})& =& (\mathbf{w}_{7}, \overline{(1,1)}) + (\mathbf{w}_{6}, \overline{(3,1)})~,
\end{eqnarray}
we see that each element on the right-hand side has the same spin.  These lines generate an Ising fusion category symmetry:
\begin{equation}
    2 \mathbf{w}_{6}\times 2 \mathbf{w}_{6}=0~, \hspace{.15in}2 \mathbf{w}_{6}\times \mathbf{w}_{7}=\mathbf{w}_{7}~,\hspace{.15in}\mathbf{w}_{7} \times \mathbf{w}_{7}=0+2 \mathbf{w}_{6}~.
\end{equation}
More precisely, keeping track of the spins of the lines, the symmetry is $Spin(5)_{1}.$

Now we make use of \eqref{E72alfacoset} to check this conclusion.  First, we must deduce the algebra $\mathcal{A}.$ There are three candidates that are consistent with the fusion rules and that saturate the quantum dimension condition \eqref{dihedral} which reads:
\begin{equation}
    \mathrm{dim}(E_{7,2}) = \frac{\mathrm{dim}(\frac{(SU(2)_{2} \times Spin(12)_{2}}{\mathbb{Z}_{2}} \times \mathrm{TetraIsing})}{\mathrm{dim}(\mathcal{A})^{2}}~.
\end{equation}
Namely: 
\begin{align}
    \mathcal{A} &= ((0,0),(1,1)) + ((0,2\mathbf{w}_{1}),(4,1)) + ((0,\mathbf{w}_{4}),(4,3))\nonumber \\[0.3cm] & + ((1,\mathbf{w}_{1} + \mathbf{w}_{s})_{2},(4,2)) + ((1,\mathbf{w}_{c})_{1},(4,4)) ~,
\end{align}
\begin{align}
    \mathcal{A}' &= ((0,0),(1,1)) + ((0,2\mathbf{w}_{1}),(4,1))+ ((0,\mathbf{w}_{4}),(4,3)) \nonumber \\[0.3cm] & + ((1,\mathbf{w}_{1} + \mathbf{w}_{s})_{1},(4,2))  + ((1,\mathbf{w}_{c})_{2},(4,4))~,
\end{align}
and
\begin{align}\label{primeprime}
    \mathcal{A}'' &= ((0,0),(1,1)) + ((0,0),(4,1)) + ((0,2\mathbf{w}_{1}),(1,1))  \nonumber \\[0.3cm] & + ((0,2\mathbf{w}_{1}),(4,1)) + 2((0,\mathbf{w}_{4}),(4,3))~.
\end{align}

To resolve this ambiguity, we recall from \eqref{condense} that we seek to present the Higgsed phase $H_{\tilde{k}}$ without 
the additional condensation $\mathcal{C}$ described in footnote \ref{footinversion}.  According to \cite{Frohlich:2003hm, Cordova:2023jip} this requires that the only anyon in the algebra $\mathcal{A}$ of \eqref{inversion} of the form $(x,1)$ has $x=1$.  Inspecting  
$\mathcal{A}''$ in \eqref{primeprime} we see the component $ ((0,2\mathbf{w}_{1}),(1,1))$ in $\mathcal{A}''$ violating this condition.  Thus, we do not consider gauging $\mathcal{A}''.$

Meanwhile, the difference between $\mathcal{A}$ and $\mathcal{A}'$ is immaterial since they differ only in which choice of split lines in $(SU(2)_{2} \times Spin(12)_{2})/\mathbb{Z}_{2}$ appear in the algebra.  We therefore proceed with the algebra object $\mathcal{A}.$

In this case, we claim that the result of the condensation again produces topological lines which form an Ising fusion category.  Indeed, the non-trivial lines in $(SU(2)_{2} \times Spin(12)_{2})/\mathbb{Z}_{2}$ that remain topological are $(0,2\mathbf{w}_{s})$ and $(1,\mathbf{w}_{3})_{2}$. As a check, we can calculate the fusion with the algebra and see that the uniform spin condition \eqref{fusionwithalgebra} is fulfilled:
\begin{eqnarray}
   \mathcal{A} \times ((0,2\mathbf{w}_{s}), (1,1)) &=&  ((0,2\mathbf{w}_{s}), (1,1))  + ((2,0), (4,1)) \nonumber\\
& +& ((0,\mathbf{w}_{2}), (4,3))  + ((1,\mathbf{w}_{c})_{2}, (4,2)) \nonumber\\
&+& ((1,\mathbf{w}_{1} + \mathbf{w}_{s})_{1}, (4,4))~, 
\end{eqnarray}
and
\begin{eqnarray}
    \mathcal{A} \times ((1,\mathbf{w}_{3})_{2}, (1,1)) &=& ((1,\mathbf{w}_{3})_{2}, (1,1))   + ((1,\mathbf{w}_{3})_{1}, (4,1))\nonumber\\ 
    &+& ((1,\mathbf{w}_{1}), (4,3))  + ((0,\mathbf{w}_{s}), (4,2))\nonumber\\ 
    &+& ((0,\mathbf{w}_{1} + \mathbf{w}_{c}), (4,4))~.
\end{eqnarray}
The fusions of these lines are:
\begin{equation}
    (0,2\mathbf{w}_{s}) \times (0,2\mathbf{w}_{s}) = (0,0)~,\hspace{.1in}(0,2\mathbf{w}_{s}) \times (1,\mathbf{w}_{3})_{2} = (1,\mathbf{w}_{3})_{2}~,
\end{equation}
as well as:
\begin{equation}
(1,\mathbf{w}_{3})_{2} \times (1,\mathbf{w}_{3})_{2} = (0,0) + (0,2\mathbf{w}_{s})~,
\end{equation}
 which again define an Ising fusion ring $Spin(5)_{1}$. \\



\let\oldaddcontentsline\addcontentsline
\renewcommand{\addcontentsline}[3]{}
\section*{Acknowledgements}
\let\addcontentsline\oldaddcontentsline

We thank M. Levin, J. McNamara, and J. McGreevy for discussions. CC, and DGS acknowledge support from the US Department of Energy Grant 5-29073, and the Sloan Foundation. KO is supported by JSPS KAKENHI Grant-in-Aid No.22K13969 and No.24K00522. CC, DGS, and KO also acknowledge support by the Simons Foundation Grant \#888984 (Simons Collaboration on Global Categorical Symmetries). Some figures were created using the Makie package \cite{DanischKrumbiegel2021}.

\appendix

\section{A Quantum Dimension Check} \label{AppendixA}

In this appendix we briefly check that the quantum dimension constraint on the algebra $\mathcal{D}_{k}$ mentioned in Section \ref{sec:confembed} is indeed fulfilled. Recall that we wish to check that the algebra is such that
\begin{equation} \label{qdimconstraint}
    \frac{\mathrm{dim}(SU(k)_{2} \times SU(2)_{k})}{\mathrm{dim}(\mathcal{D}_{k})^{2}} = \mathrm{dim}(SU(2k)_{1})~,
\end{equation}
where
\begin{equation}
    \mathrm{dim}(SU(k)_{2}) = \frac{k(k+2)}{4 \sin{(\frac{\pi}{k+2})}^{2}}~, 
\end{equation}
\begin{equation}
    \mathrm{dim}(SU(2)_{k}) = \frac{(k+2)}{2 \sin{(\frac{\pi}{k+2})}^{2}}~,
\end{equation}
and
\begin{equation}
    \mathrm{dim}(SU(2k)_{1}) = 2k~.
\end{equation}
Indeed, recall that the quantum dimensions of the $SU(2)_{k}$ fusion ring are
\begin{equation}
    d^{SU(2)_{k}}_{j} = \frac{\sin{(\frac{(j+1)\pi}{k+2})}}{\sin{(\frac{\pi}{k+2})}}~,
\end{equation}
where $j=0,1,\ldots,k-1$. Recall as well that the algebra is composed by the diagonal subset of lines inside the $PSU(2)_{-k} \times PSU(2)_{k}$ subfusion ring of $SU(k)_{2} \times SU(2)_{k}$. Then, the dimension of the algebra is
\begin{eqnarray}
    \mathrm{dim}(\mathcal{D}_{k}) = \sum_{j \ \mathrm{even}} (d^{SU(2)_{k}}_{j})^{2}~.
\end{eqnarray}
This is easily calculated, and indeed \eqref{qdimconstraint} is obeyed.


\let\oldaddcontentsline\addcontentsline
\renewcommand{\addcontentsline}[3]{}
\bibliography{references}
\let\addcontentsline\oldaddcontentsline

\end{document}